\documentclass[english,aps,prc,nofootinbib,superscriptaddress,twocolumn,letter]{revtex4}
\usepackage[latin9]{inputenc}
\usepackage{hyperref}
\hypersetup{
  colorlinks=true,        
  linkcolor=blue,         
  citecolor=cyan,         
}
\usepackage{breakurl}
\usepackage{graphicx}
\usepackage{epsf}
\usepackage{epsfig}
\usepackage{amssymb,amsmath}
\usepackage[usenames]{color}
\usepackage{amssymb}
\usepackage{times}
\usepackage{comment}

\newcommand\pt{p_\text{T}}

\allowdisplaybreaks



\begin{document}

\preprint{This line only printed with preprint option}

\title{Thermal photon emission from quark-gluon plasma: 1+1D magnetohydrodynamics results}

\author{Jie Xiong}
\affiliation{Department of Physics and Electronic-Information Engineering, Hubei Engineering University, Xiaogan, Hubei, 432000, China}

\author{Xiang Fan}
\affiliation{Institute of Particle Physics and Key Laboratory of Quark and Lepton Physics (MOE), Central China Normal University, Wuhan, Hubei, 430079, China}

\author{Jing Jing}
\affiliation{Department of Physics and Electronic-Information Engineering, Hubei Engineering University, Xiaogan, Hubei, 432000, China}

\author{Weishan Yang}
\affiliation{Department of Physics and Electronic-Information Engineering, Hubei Engineering University, Xiaogan, Hubei, 432000, China}

\author{Duan She}
\affiliation{Institute of Physics, Henan Academy of Sciences, Zhengzhou 450046, China}

\author{Ze-Fang Jiang}
\email{jiangzf@mails.ccnu.edu.cn}
\affiliation{Department of Physics and Electronic-Information Engineering, Hubei Engineering University, Xiaogan, Hubei, 432000, China}
\affiliation{Institute of Particle Physics and Key Laboratory of Quark and Lepton Physics (MOE), Central China Normal University, Wuhan, Hubei, 430079, China}

\begin{abstract}
We investigate thermal photon production in the quark-gluon plasma (QGP) under strong magnetic fields using a magnetohydrodynamic (MHD) framework. Adopting the Bjorken flow model with power-law decaying magnetic fields $\mathbf{B}(\tau) = \mathbf{B}_0 (\tau_0/\tau)^a$ (where $a$ controls the decay rate, $B_0 = \sqrt{\sigma} T_0^2$, and $\sigma$ characterizes the initial field strength), we employ relativistic ideal fluid dynamics under the non-resistive approximation. The resulting QGP temperature evolution exhibits distinct $a$- and $\sigma$-dependent behaviors. Thermal photon production rates are calculated for three dominant processes: Compton scattering with $q\bar{q}$ annihilation (C+A), bremsstrahlung (Brems), and $q\bar{q}$ annihilation with additional scattering (A+S). These rates are integrated over the space-time volume to obtain the photon transverse momentum $(p_T)$ spectrum. 
Our results demonstrate that increasing $a$ enhances photon yields across all $p_T$, with $a \to \infty$ (super-fast decay) providing an upper bound. For $a = 2/3$, larger $\sigma$ suppresses yields through accelerated cooling, whereas for $a \to \infty$, larger $\sigma$ enhances yields via prolonged thermal emission. Low-$p_T$ photons receive significant contributions from all QGP evolution stages, while high-$p_T$ photons originate predominantly from early times. The central rapidity region $(y=0)$ dominates the total yield. This work extends photon yield studies to the MHD regime under strong magnetic fields, elucidating magnetic field effects on QGP electromagnetic signatures and establishing foundations for future investigations of magnetization and dissipative phenomena.
\end{abstract}

\pacs{12.38.Mh,25.75.-q,24.85.+p,25.75.Nq}

\maketitle

\section{Introduction}

Heavy-ion collisions at RHIC and LHC provide a unique platform to study the quark-gluon plasma (QGP), a deconfined state of nuclear matter created under extreme temperature and pressure conditions~\cite{STAR:2005gfr,ALICE:2008ngc}. A striking feature of these collisions is the generation of extremely strong magnetic fields, with magnitudes ranging from \(10^{18}\) to \(10^{20}\) Gauss, induced by the rapid motion of charged spectator nucleons~\cite {Deng:2012pc,Li:2016tel,Gursoy:2014aka}. These strong magnetic fields are expected to modulate QGP dynamics through various mechanisms~\cite {Das:2016cwd,Gursoy:2018yai,Chatterjee:2018lsx,Oliva:2020doe,Sun:2020wkg,Jiang:2022uoe,Huang:2017tsq}, including quantum anomaly-induced phenomena such as the chiral magnetic effect (CME), chiral separation effect (CSE), and chiral magnetic wave (CMW)~\cite {Kharzeev:2007jp,Fukushima:2008xe,Kharzeev:2015znc,Kharzeev:2024zzm,Liang:2020sgr,Jiang:2016wve,Choudhury:2021jwd,Tang:2019pbl,Gao:2020vbh,Shi:2017cpu,Huang:2013iia,Pu:2014cwa,Kharzeev:2010gd}. Despite significant experimental efforts to detect these effects at RHIC and LHC~\cite {STAR:2015wza,CMS:2017lrw,STAR:2021mii,STAR:2021pwb,STAR:2019clv,ALICE:2019sgg,STAR:2023jdd}, extracting their signatures from collective flow backgrounds remains a critical challenge.

Relativistic hydrodynamics has emerged as a powerful tool to describe QGP evolution, successfully interpreting experimental observables such as harmonic flows and global polarization in non-central collisions~\cite{Heinz:2013th,Gale:2013da,Karpenko:2013wva,Jeon:2015dfa,Becattini:2017gcx,Becattini:2020ngo,Ze-Fang:2017ppe,Zhao:2022ayk,Zhao:2020wcd,Jiang:2021ajc,Wu:2021fjf,Jiang:2024ekh}. The dynamic interplay between electromagnetic fields and QGP requires solving full (3+1)-dimensional relativistic magnetohydrodynamic (MHD) equations~\cite{Inghirami:2016iru,Nakamura:2022idq,Mayer:2024kkv}. Lattice QCD calculations confirm that the QGP exhibits a temperature-dependent electrical conductivity $\sigma_{el}$ ~\cite{Ding:2010ga,Ding:2016hua}, yet the interaction between initial magnetic fields and the QGP medium~\cite{Bali:2013owa,Pang:2016yuh,Jiang:2024bez,Huang:2024aob}, along with the field's subsequent evolution, remains incompletely understood. 
Recent advances have explored electromagnetic effects within hydrodynamic frameworks~\cite{Fukushima:2024tkz}, including 1+1 dimensional Bjorken flow in ideal MHD-where energy density evolves via the ``frozen-flux theorem''~\cite{Roy:2015kma,Pu:2016ayh,Siddique:2019gqh,Peng:2022cya}-and extensions incorporating magnetization~\cite{Pu:2016ayh}, longitudinal expansion~\cite{She:2019wdt,HaddadiMoghaddam:2020ihi}, and rotating solutions~\cite{Shokri:2018qcu}. Concurrently, efforts to address causality and stability in relativistic dissipative MHD, particularly through the Israel-Stewart formalism, have advanced understanding of non-resistive magnetohydrodynamic behaviors~\cite{Biswas:2020rps,Cordeiro:2023ljz,Fang:2024skm,Rocha:2023ilf,Most:2021rhr}.

Thermal photons (and dileptons) serve as one of the cleanest probes of QGP properties~\cite{Bhatt:2010cy,Gale:2003iz,Gale:2009gc,dEnterria:2005jvq,Linnyk:2013hta,Wang:2020dsr,Steffen:2001pv,Naik:2025pjt,Wang:2021eud}, emitted throughout all stages of heavy-ion collisions and escaping the medium without significant reabsorption. Originating from initial hard scatterings, medium-induced thermal radiation, and late-stage hadronic decays, they carry useful information about the QGP's thermodynamic and dynamical history~\cite{Steffen:2001pv,Shen:2013cca,David:2019wpt,Schenke:2006yp,Wu:2024vyc,Chatterjee:2017akg}. Recent studies highlight their utility in probing initial states of the fireball~\cite{Vovchenko:2016ijt}, measuring transport coefficients like shear/bulk viscosity, and extend the knowledge of the differential emission rates from a strongly magnetized plasma~\cite{Wang:2021eud}. Thermal photon spectra depend sensitively on the QGP temperature evolution, which is governed by relativistic hydrodynamics with appropriate initial conditions. Key production mechanisms include quark-antiquark annihilation (\(q\bar{q} \to g\gamma\)), Compton scattering (\(q(\bar{q})g \to q(\bar{q})\gamma\)), bremsstrahlung processes, and soft processes calculated within hard thermal loop perturbation theory~\cite{Traxler:1995kx,Steffen:2001pv,Bhatt:2010cy,Kasmaei:2019ofu,David:2019wpt,Schenke:2006yp}, whose contributions vary across momentum ranges and encode details of the medium's electromagnetic response. 

Against this backdrop, the present work investigates thermal photon production in the QGP stage under the influence of external magnetic fields within an ideal magnetohydrodynamic framework. Building on Victor-Bjorken flow assumptions~\cite{Roy:2015kma,Pu:2016ayh,Jiang:2024mts}, we analyze how magnetic field dynamics-characterized by a power-law decay \(\overrightarrow{B}(\tau) = \overrightarrow{B}_0 (\tau_0/\tau)^a\) with decay parameter \(a\) and initial strength parameter \(\sigma\)-modulate the QGP temperature evolution and subsequent photon emission. We calculate photon production rates for three dominant processes: Compton scattering with \(q\bar{q}\) annihilation (C+A), bremsstrahlung (Bre), and \(q\bar{q}\) annihilation with additional scattering (A+S), integrating these rates over the QGP's spacetime history to obtain observable spectra. By systematically varying \(a\) and \(\sigma\), we quantify how magnetic field decay kinetics and initial strength influence photon yields across transverse momentum ranges, rapidity intervals, and temporal stages of QGP evolution. This study not only extends previous analyses of photon production in viscous and accelerated fluids to the magnetohydrodynamic~\cite{Kasza:2025wot} regime but also establishes a foundational framework for future investigations into magnetization, dissipative effects~\cite{Bhatt:2010cy}, and spin-dependent phenomena~\cite{Wang:2024gnh,Wu:2024vyc} in relativistic heavy-ion collisions. 

This paper is organized as follows. In Sec.~\ref{sec:2}, we introduce the MHD framework and photon rate formalisms. In Sec.~\ref{section-4}, we present results for photon production. Finally, we summarize in the Sec.~\ref{section-5}. Throughout this work, $u^{\mu}=\gamma\left(1,\overrightarrow{\boldsymbol{v}}\right)$ is the four-velocity field that satisfies $u^{\mu}u_{\mu}=1$ and the spatial projection operator $\Delta^{\mu\nu}=g^{\mu\nu}-u^{\mu}u^{\nu}$ is defined with the Minkowski metric $g^{\mu\nu}={\rm diag}\left(1,-1,-1,-1\right)$. It is note-worthy that the orthogonality relation $\Delta^{\mu\nu}u_{\nu}=0$ is satisfied.

\section{Formalism}
\label{sec:2}

\subsection{Magnetohydrodynamic Framework}
\label{sec:2-A}

The total energy-momentum tensor of relativistic fluids in the presence of magnetic fields is described by the causal second-order Israel-Stewart (IS) theory-a key framework for capturing dissipative effects in relativistic systems while ensuring causality and stability~\cite{Denicol:2018rbw,Biswas:2020rps}:
\begin{equation}
\begin{aligned}
T^{\mu\nu} &= (\varepsilon + p + \Pi + E^{2}+ B^{2})u^{\mu}u^{\nu} \\
           &-\left(p+ \Pi+\frac{1}{2}E^{2}+\frac{1}{2}B^{2}\right)g^{\mu\nu}\\
&-E^{\mu}E^{\nu}-B^{\mu}B^{\nu}-u^{\mu}\epsilon^{\nu\lambda\alpha\beta}E_{\lambda}B_{\alpha}u_{\beta} \\
& -u^{\nu}\epsilon^{\mu\lambda\alpha\beta}E_{\lambda}B_{\alpha}u_{\beta} + \pi^{\mu\nu},
\label{tmunu_total}
\end{aligned}
\end{equation}
where $\varepsilon$ and $p$ denote the energy density and pressure, $u^{\mu}$ is the fluid four-velocity, $\pi^{\mu\nu}$ the shear viscous tensor, and $\Pi$ the bulk viscous pressure. The magnetic ($B^\mu$) and electric ($E^\mu$) field four-vectors are defined as
\begin{equation}
B^{\mu} = \frac{1}{2}\epsilon^{\mu\nu\alpha\beta}u_{\nu}F_{\alpha\beta}, \quad E^{\mu}=F^{\mu\nu}u_{\nu},
\end{equation}
with $F^{\mu\nu}=\partial^\mu A^\nu - \partial^\nu A^\mu$ (Faraday tensor, encoding electromagnetic field dynamics) and $\epsilon^{\mu\nu\alpha\beta}$ (Levi-Civita tensor, with $\epsilon^{0123}=+1$ for orientation). These fields satisfy $u^\mu E_\mu = u^\mu B_\mu = 0$, confirming their spacelike nature, and $B^\mu B_\mu = -B^2$, $E^\mu E_\mu = -E^2$ (where $B$ and $E$ are field magnitudes).

In the non-resistive limit (infinite electrical conductivity $\sigma_{el}$), the electric field $E^\mu \to 0$ to ensure finite charge current $j^\mu = \sigma_{el} E^\mu$, a key constraint for ideal magnetohydrodynamics (MHD)~\cite{Roy:2015kma}. Here, magnetic field evolution is governed by $\partial_\nu (B^\mu u^\nu - B^\nu u^\mu) = 0$, reflecting the "frozen-flux theorem" where field lines are advected with the fluid. For ideal fluids (neglecting viscosity, $\Pi=\pi^{\mu\nu}=0$), the energy-momentum tensor simplifies to~\cite{Roy:2015kma,Biswas:2020rps}:
\begin{equation}
T^{\mu\nu} = (\varepsilon + p + B^{2})u^{\mu}u^{\nu}-\left(p+\frac{1}{2}B^{2}\right)g^{\mu\nu}-B^{\mu}B^{\nu}.
\label{tmunu_t}
\end{equation}

The system evolves via energy-momentum conservation:
\begin{equation}
\partial_{\mu}T^{\mu\nu}=0.
\end{equation}
To close the equations, we adopt a conformal equation of state (EoS) for high-temperature QGP: $\varepsilon = 3p$ (implying sound speed $c_s^2 = 1/3$). This approximation is valid for the deconfined phase where quark-gluon degrees of freedom dominate, though we note that lattice QCD calculations predict a temperature-dependent $c_s(T)$ for a more realistic description. Additionally, we neglect QGP magnetization~\cite{Pu:2016ayh}, assuming isotropic pressure and no magnetic-field-induced modifications to the EoS.

Under longitudinal boost invariance-appropriate for high-energy heavy-ion collisions where the fireball expands primarily along the beam axis-we use Milne coordinates: $t = \tau \cosh\eta_s$, $z = \tau \sinh\eta_s$. Here, $\tau = \sqrt{t^2 - z^2}$ is the proper time (invariant under longitudinal boosts), and $\eta_s = 0.5\ln[(t+z)/(t-z)]$ is the space-time rapidity. The fluid four-velocity takes the form
\begin{equation}
u^{\mu} = (\cosh\eta_s, 0, 0, \sinh\eta_s) = \gamma(1, 0, 0, z/t),
\end{equation}
with $\gamma = \cosh\eta_s$ as the Lorentz factor, capturing relativistic effects in the longitudinal direction.

Following Refs.~\cite{Roy:2015kma,Pu:2016ayh,She:2019wdt,Biswas:2020rps}, the magnetic field is assumed to decay via a power law in proper time:
\begin{equation}
\overrightarrow{B}(\tau) = \overrightarrow{B}_0 \left(\frac{\tau_0}{\tau}\right)^a,
\label{11}
\end{equation}
where $a > 0$ is the decay constant (controlling decay rate), $\tau_0$ is the initial proper time, and $B_0 \equiv B(\tau_0)$ is the initial field strength.

Projecting $\partial_\mu T^{\mu\nu} = 0$ onto the fluid four-velocity $u_\mu$ (Landau-Lifshitz frame, where energy flow is measured in the fluid rest frame) gives the energy conservation equation:
\begin{equation}
\frac{\partial \left(\varepsilon + \frac{1}{2}B^{2}\right)}{\partial \tau} + \frac{\varepsilon + p + B^{2}}{\tau} = 0,
\label{energy_conservation_1}
\end{equation}
with $u \cdot \partial = \partial/\partial\tau$ (time derivative in the fluid frame), $\partial_\mu u^\mu = 1/\tau$ (expansion factor, reflecting longitudinal dilution), and $\nabla^\mu = \Delta^{\mu\nu}\partial_\nu$ (spatial gradient operator)~\cite{Muronga:2001zk,Muronga:2003ta}.

Projecting orthogonally to $u^\mu$ via 
\begin{equation}
\Delta_{\mu\nu}\partial_{\alpha}T^{\alpha\nu} = 0
\end{equation}
yields the momentum conservation equation:
\begin{equation}
\nabla^{\mu}\left(p + \frac{B^{2}}{2}\right) - (\varepsilon + p + B^{2})\frac{\partial u^{\mu}}{\partial\tau} = 0.
\label{euler}
\end{equation}
For the space-time rapidity component $\mu = \eta_s$, this simplifies to 
\begin{equation}
\frac{\partial}{\partial\eta_{s}}\left(p+\frac{1}{2} B^{2}\right)=0,
\end{equation}
implying thermodynamic variables are uniform in $\eta_s$ (spatially homogeneous) and depend solely on $\tau$~\cite{Muronga:2003ta,Roy:2015kma}. Transverse velocities (in $x$,$y$ directions) remain zero due to initial symmetry, as no net forces arise to induce motion.

For simplicity, we use following thermodynamic relations~\cite{Roy:2015kma,Muronga:2003ta,Pu:2016ayh}:
\begin{equation}
\begin{aligned}
& p = a_1 T^4, \quad p = c_s^2 \varepsilon = \varepsilon/3, \\
& B_0^2 = \sigma_{0}\varepsilon_{0}=3\sigma_{0}p_{0}=\sigma T_0^4,
\label{e_simp}
\end{aligned}
\end{equation}
where $T_0$ is the initial temperature at $\tau_0$, and
\begin{equation}
a_1 = \left(16 + \frac{21}{2}N_f\right)\frac{\pi^2}{90}
\label{rt_t}
\end{equation}
is a constant determined by the number of quark flavors $N_f$ and gluon colors. The dimensionless parameter $\sigma$ characterizes initial magnetic field strength. For RHIC energies ($T_0 = 0.31$ GeV) with $B_0 \in [10^{18}, 10^{19}]$ Gauss (generated by rapid motion of charged spectator nucleons), results in $\sigma$ ranges from $0.04$ to $40$. Notably, \(B_0^2 = \sigma T_0^4\) is an effective estimation of the initial magnetic field strength based on vacuum calculations for relativistic heavy-ion collisions [{\color{cyan}55,56}], valid for central-to-moderately non-central collisions in the ideal MHD limit. This relation fails for finite QGP conductivity, magnetized QGP, or ultra-peripheral collisions, where the initial field depends strongly on the impact parameter \(b\) and requires spatial-resolved calculations.

\begin{figure}[tbp!]
\includegraphics[width=0.85\linewidth]{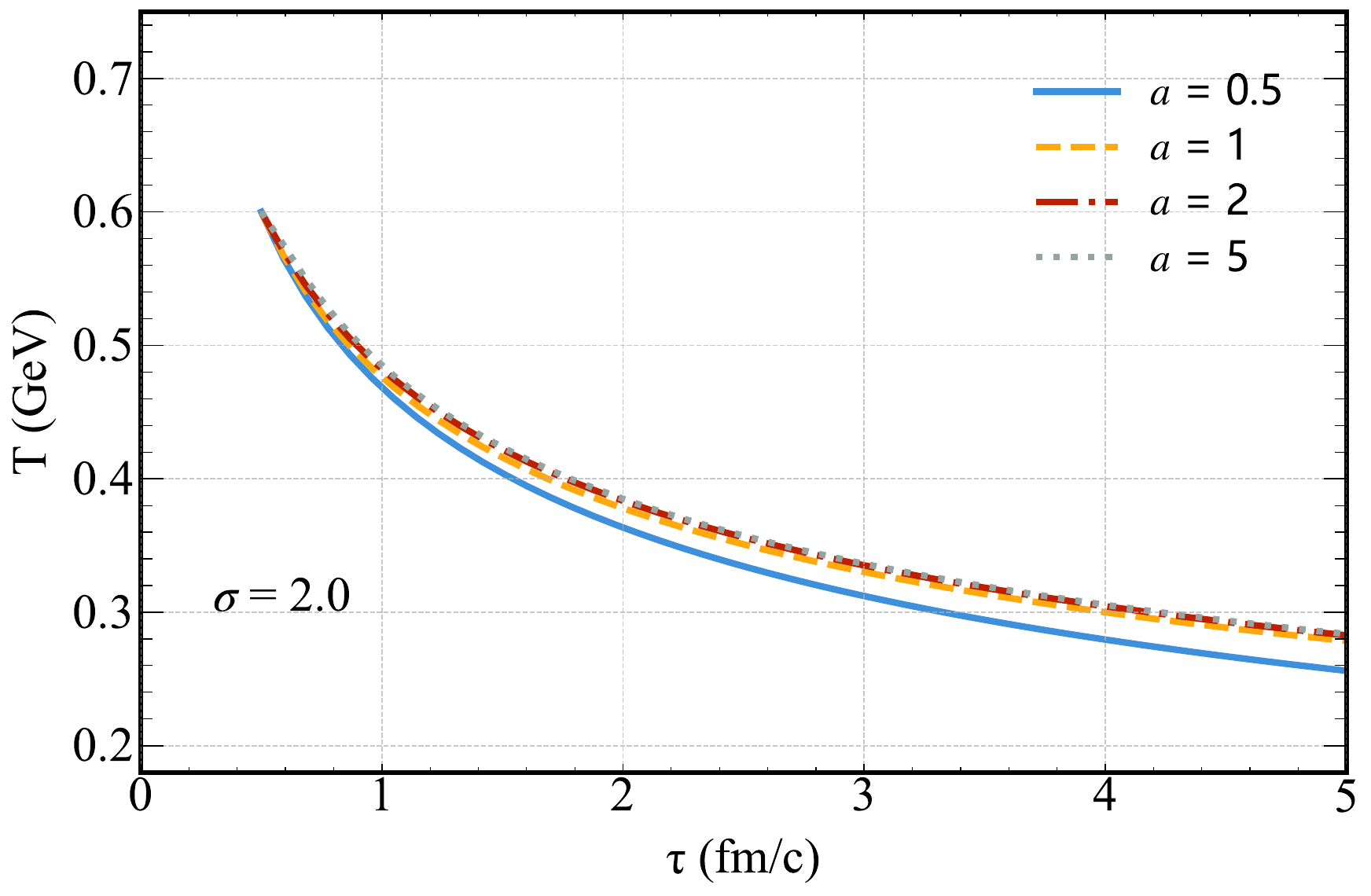} \\
\includegraphics[width=0.85\linewidth]{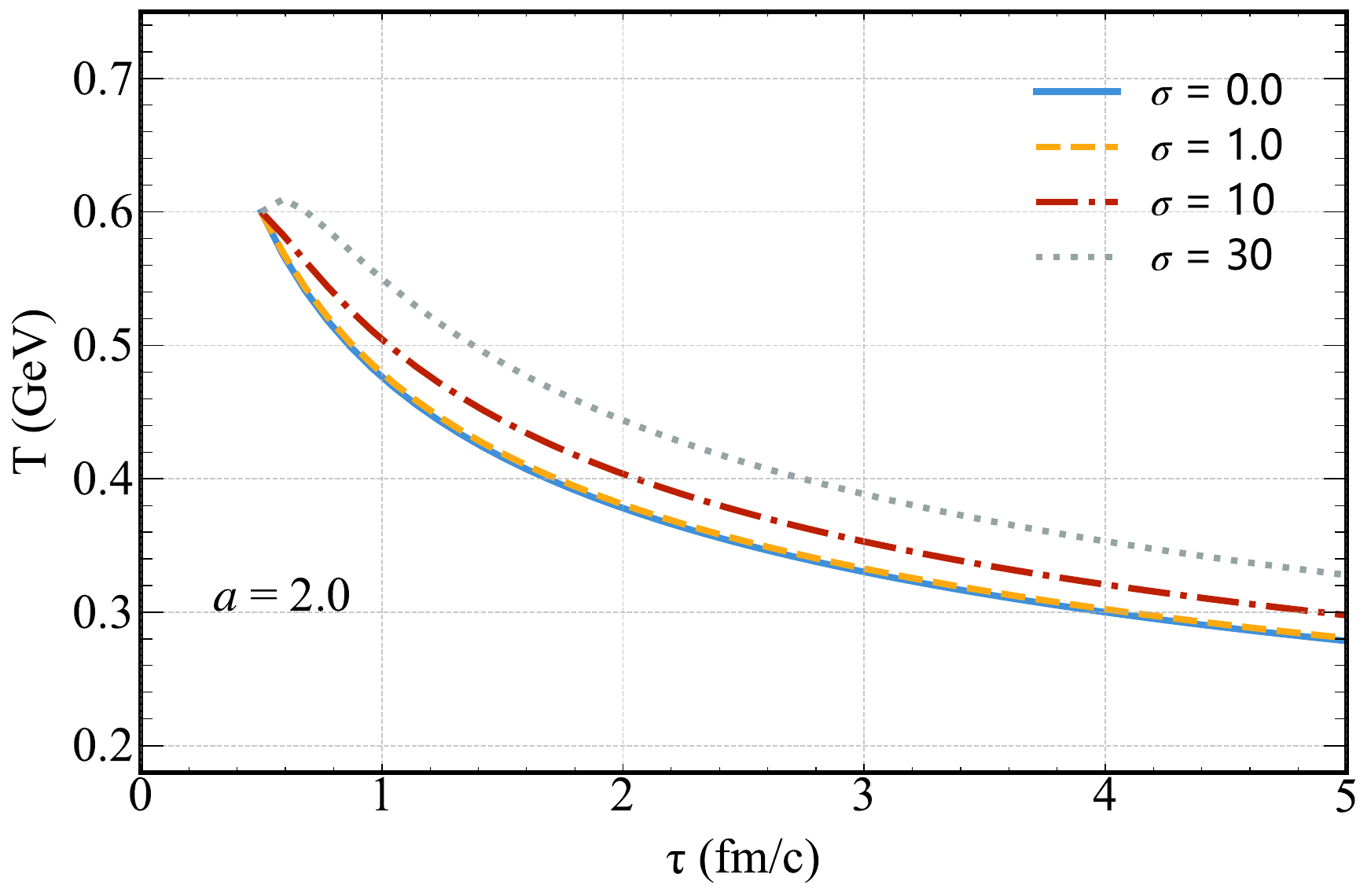}
\caption{(Color online) Evolution of temperature $T$ (Eq.~(\ref{T_mhd_1})) for a magnetic field with as functions of proper time $\tau$ for different initial magnetic field (upper panel) and different magnetic field decay parameter $a$ (lower panel).}
\label{f:ideal-mhd}
\end{figure}

\subsection{Analytical Solution for the MHD Flow}
\label{sec:2-B}

We begin by analyzing the 1+1 dimensional ideal magnetohydrodynamic (MHD) flow-known as the \textit{Victor-Bjorken} flow~\cite{Roy:2015kma}-a model for describing longitudinal expansion in high-energy heavy-ion collisions. This framework simplifies the problem by assuming boost invariance along the beam axis, making it tractable for exploring magnetic field effects on QGP evolution. Starting from the energy conservation equation (Eq.~(\ref{energy_conservation_1})), we incorporate contributions from the external magnetic field and utilize the thermodynamic relations from Eq.~(\ref{e_simp}) (e.g., $p = a_1 T^4$ and $B_0^2 = \sigma T_0^4$) to derive a temperature-dependent form of the conservation law. This reduces to:

\begin{equation}
\begin{aligned}
\frac{\partial T}{\partial \tau} + \frac{(1-a) \sigma T_{0}^{4}}{12 a_{1} T^{3}}\left(\frac{\tau_{0}}{\tau}\right)^{2a}\frac{1}{\tau} + \frac{T}{3\tau} = 0.
\label{MHD_energy_1}
\end{aligned}
\end{equation}

Here, the first term represents the local rate of temperature change, the second term captures the magnetic field's influence (via $\sigma$ and decay parameter $a$), and the third term accounts for longitudinal expansion (consistent with Bjorken scaling). The analytical solution to this equation is:

\begin{equation}
\begin{aligned}
T=T_{0}\left(\frac{\tau_{0}}{\tau}\right)^{\frac{1}{3}}\left[1+\frac{\sigma(a-1)}{2a_{1}(3a-2)}\left(1-\left(\frac{\tau_{0}}{\tau}\right)^{2a-\frac{4}{3}}\right)\right]^{\frac{1}{4}}.
\label{T_mhd_1}
\end{aligned}
\end{equation}
This solution combines two key components: the standard Bjorken-like decay $\propto (\tau_0/\tau)^{1/3}$ and a magnetic field correction term, which modifies the temperature based on $\sigma$ and $a$.

The upper panel of Fig.~\ref{f:ideal-mhd} illustrates the temperature $T$ as a function of proper time $\tau$ for $\sigma=2$, with varying magnetic field decay parameters $a=0.5,~1,~2,~5$ (initial conditions: $\tau_0=0.5$ fm/c, $T_0=0.6$ GeV). For $a>0$, a clear trend emerges: the normalized temperature decays more slowly as $a$ increases. This behavior arises because faster magnetic field decay (larger $a$) transfers energy from the field to the QGP more efficiently, a process described as ``reheating'' of the energy density~\cite{Roy:2015kma}. For example, $a=5$ (rapid decay) retains higher temperatures than $a=0.5$ (slow decay) at late times, as the field's energy is deposited into the medium before significant expansion cools it.

The lower panel of Fig.~\ref{f:ideal-mhd} plots $T(\tau)$ for a fixed decay parameter $a=2$ and varying initial magnetic field strengths $\sigma=0.01,~1,~10,~30$. A larger $\sigma$ (stronger initial field) slows the decay of $T$, with the most pronounced effect for $\sigma=30$. Notably, the magnetic correction term in Eq.~(\ref{T_mhd_1}) is non-monotonic in $\tau$, leading to an initial "reheating" spike for sufficiently large $\sigma$ (e.g., $\sigma=30$). This spike occurs because the strong initial field releases energy into the QGP faster than expansion can cool it, temporarily reversing the temperature decay.

Consistent with prior studies~\cite{Roy:2015kma,Pu:2016ayh,Jiang:2024mts}, we examine the divergent behavior of Eq.~(\ref{T_mhd_1}) under specific limits of $a$, which govern the magnetic field's decay rate and its coupling to the QGP:

\emph{\textbf{Case A}} For $a=1$ (ideal MHD limit with infinite conductivity, corresponding to maximal magnetic induction), the magnetic field is "frozen" into the fluid, and the correction term in Eq.~(\ref{T_mhd_1}) vanishes. This reduces the solution to the standard Victor-Bjorken flow~\cite{Roy:2015kma}:
\begin{equation}
\begin{aligned}
T = T_{0}\left(\frac{\tau_0}{\tau}\right)^{\frac{1}{3}}.
\label{T_mhd_2}
\end{aligned}
\end{equation}
Here, the magnetic field does not modify the temperature decay, as energy transfer between the field and fluid is balanced by expansion.

\emph{\textbf{Case B}} In the limit $a \rightarrow \frac{2}{3}$, the correction term in Eq.~(\ref{T_mhd_1}) simplifies via L'Hospital's rule, yielding a logarithmic dependence:
\begin{equation}
\begin{aligned}
\lim\limits_{a \rightarrow \frac{2}{3}}\frac{\sigma (a-1)}{2a_{1}(3a - 2)} \left(1 - \left(\frac{\tau_0}{\tau}\right)^{2a - \frac{4}{3}}\right)
=\frac{\sigma}{9 a_{1}}\ln\left(\frac{\tau_{0}}{\tau}\right).
\label{T_mhd_3}
\end{aligned}
\end{equation}
Collecting terms, the temperature evolution becomes:
\begin{equation}
\begin{aligned}
T=T_{0}\left(\frac{\tau_{0}}{\tau}\right)^{\frac{1}{3}}\left[1+\frac{\sigma}{9a_{1}}\ln\left(\frac{\tau_{0}}{\tau}\right)\right]^{\frac{1}{4}}.
\label{T_mhd_4}
\end{aligned}
\end{equation}
For $\tau>\tau_0$, the logarithmic term is negative (since $\ln(\tau_0/\tau) < 0$), reducing $T$ and accelerating its decay. This reflects a regime where magnetic field decay drains energy from the QGP faster than expansion alone, enhancing cooling.

\emph{\textbf{Case C}} For $a \rightarrow \infty$ (super-fast magnetic field decay), the field dissipates almost instantaneously, depositing its energy into the QGP early in the evolution. In this limit:
\begin{equation}
\begin{aligned}
\lim\limits _{a\rightarrow\infty}\frac{\sigma (a-1)}{2a_{1}(3a - 2)} \left(1 - \left(\frac{\tau_0}{\tau}\right)^{2a - \frac{4}{3}}\right)=\frac{\sigma}{6a_{1}}.
\end{aligned}
\end{equation}
The temperature evolution simplifies to:
\begin{equation}
\begin{aligned}
T = T_{0}\left(\frac{\tau_0}{\tau}\right)^{\frac{1}{3}} \left( 1 + \frac{\sigma}{6 a_{1}}\right)^{\frac{1}{4}}.
\label{T_mhd_5}
\end{aligned}
\end{equation}
Here, super-fast field decay slows temperature decay by adding a constant offset to the Bjorken term, establishing an upper bound on $T$ for a given $\sigma$. This behavior aligns partially with current magnetic field evolution models, particularly for $a\geq1$, where field decay efficiently reheats the QGP.

\begin{figure}[tbp!]
\includegraphics[width=0.85\linewidth]{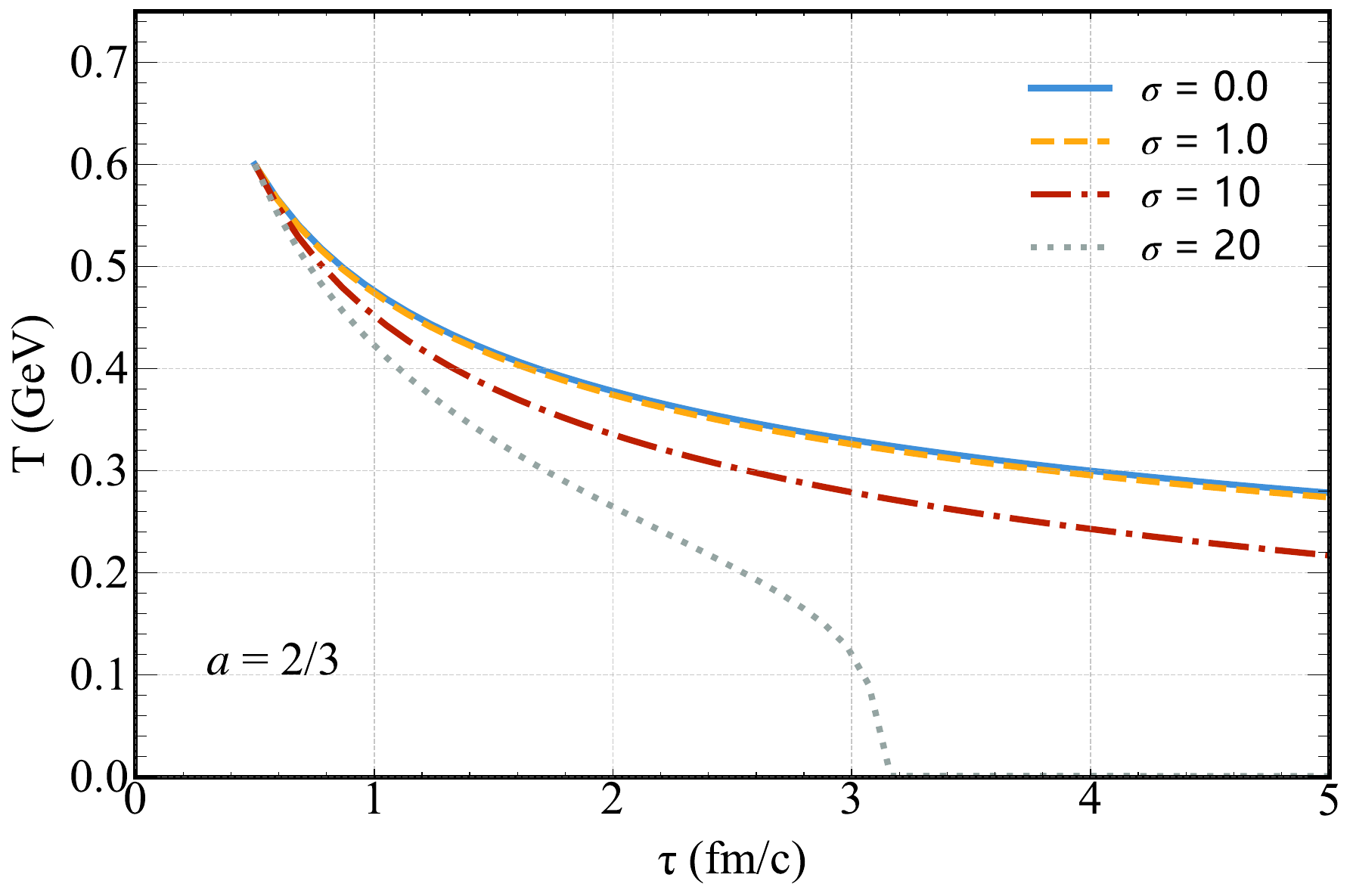}
\includegraphics[width=0.85\linewidth]{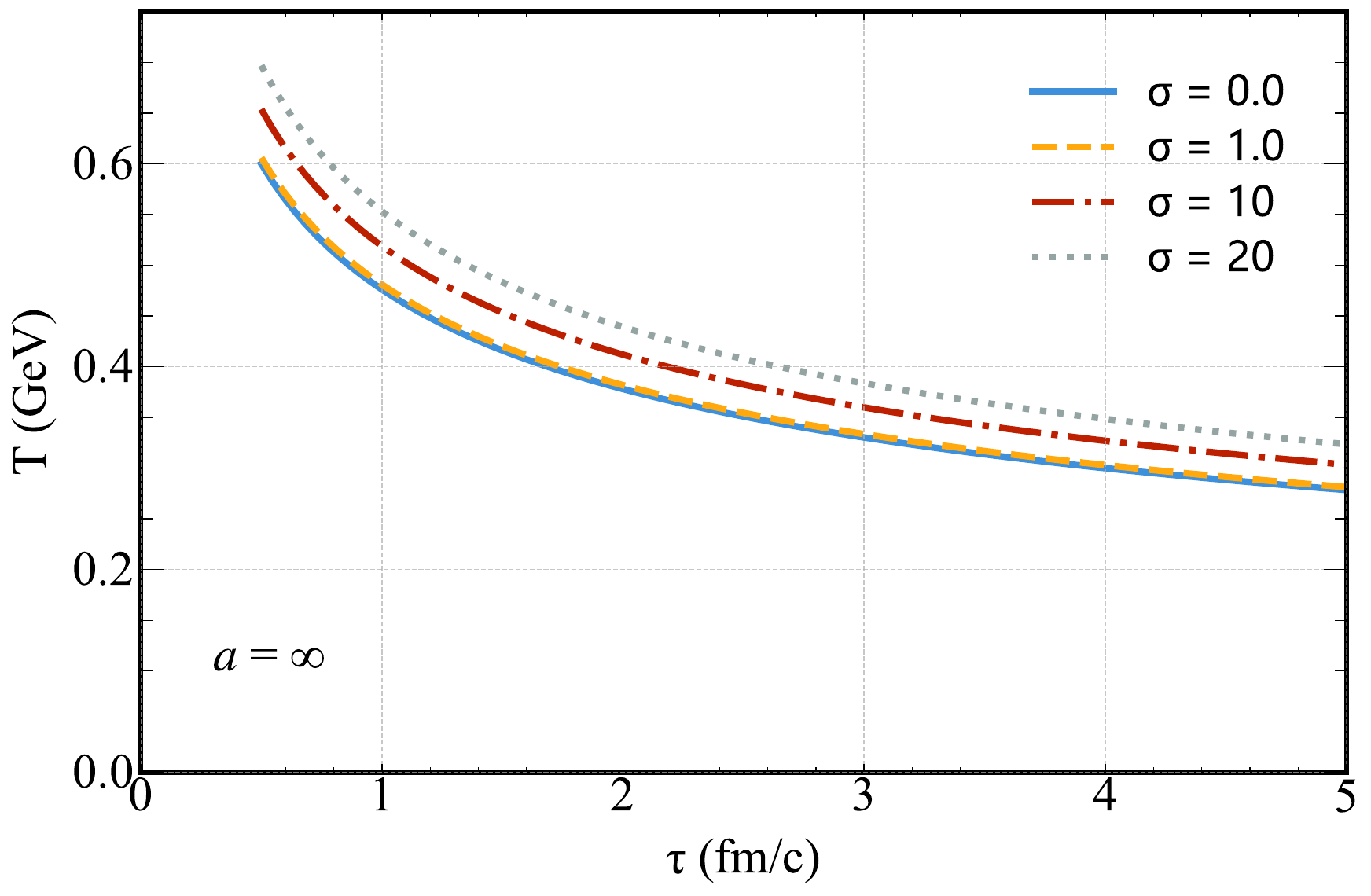}
\caption{(Color online) Evolution of temperature $T$ as functions of proper time $\tau$ for different initial magnetic field decay parameter $a$ and strength $\sigma$.}
\label{f:ns-1}
\end{figure}

In Fig.~\ref{f:ns-1} upper panel, we present the temperature $T$ as a function of proper time for the specified limit $a\rightarrow 2/3$ (Eq.~(\ref{T_mhd_4})) with various $\sigma$ = 0, 1, 10, 20. As already analysis in the previous section, for $\tau>\tau_{0}$ the log term always reduces the value of $T_{0}$, leading to a faster decrease of the temperature when compared with the ideal-MHD limit ($\sigma=0.0$) (This is shown with a solid-blue line). Furthermore, it is also clear that larger values of $\sigma$ will lead to a faster decrease in $T$ (note that for $\sigma=20$, $T\approx 0$ at $\tau=3.2$ fm). 

In Fig.~\ref{f:ns-1} lower panel, we plot the evolution of temperature $T$ in the case $a\rightarrow \infty$ (Eq.~(\ref{T_mhd_5})). Also in this case, different lines refer to different levels of the initial magnetization strength $\sigma$. Since the second term $(1+\sigma/6a_{1})$ in Eq.~(\ref{T_mhd_5}) is always positive for the  positive $\sigma$, the evolution of the temperature is expected to be slower than in standard ideal-MHD limit. 

\begin{figure}[tbp!]
\includegraphics[width=0.85\linewidth]{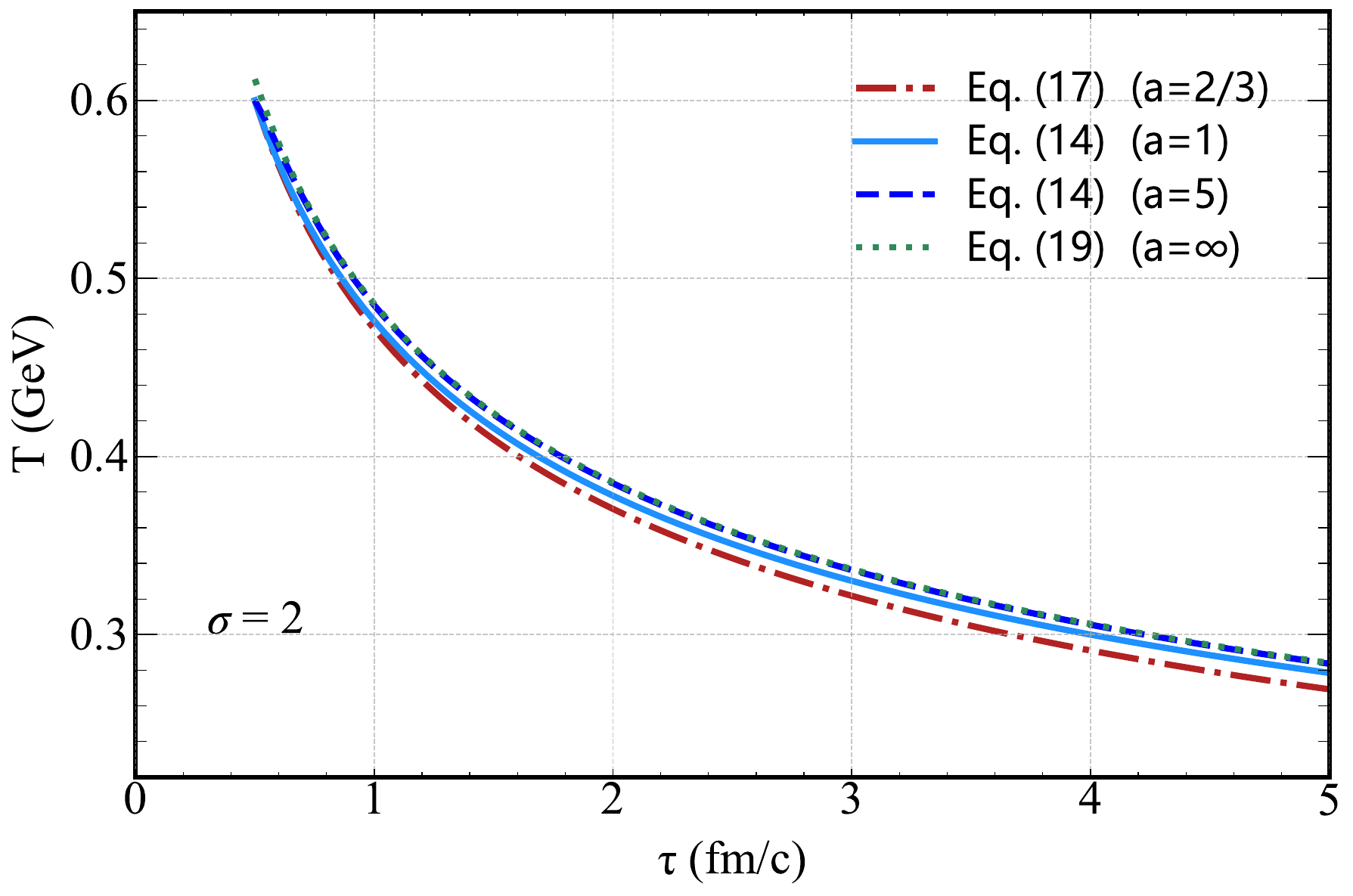} 
\caption{(Color online) Evolution of temperature $T$ as a function of proper time $\tau$ for different limit of magnetic field decay parameter $a$ with $\sigma=2$.}
\label{f:ns-2}
\end{figure}

In Fig.~\ref{f:ns-2}, we further show the evolution of temperature $T$ in the different limit cases (Eq.(\ref{T_mhd_1}), Eq.~(\ref{T_mhd_4}), Eq.~(\ref{T_mhd_5})) but keeping the initial magnetization fixed to $\sigma=2$. More specifically, we show the evolution for $a=2/3$, $a=1$, $a=5$ and $a=\infty$. Clearly, $T$ decreases more rapidly for $a=2/3$ when compared to the case $a=1$, whereas for $a=5$ it decreases more slowly and almost decays asymptotically at the same rate for the $a=\infty$ case.

\subsection{Thermal Photons}
\label{sec:2-C}

Thermal photons emitted during the quark-gluon plasma (QGP) phase serve as critical probes of the medium's thermodynamic and dynamical properties, as they escape the QGP without significant reabsorption, carrying unaltered information about their production environment. 
These photons originate from several key quantum chromodynamic (QCD) processes. The primary mechanisms include Compton scattering ($q(\bar{q})g \to q(\bar{q})\gamma$), where a quark (or antiquark) scatters off a gluon and emits a photon, and quark-antiquark annihilation ($q\bar{q} \to g\gamma$), where a quark-antiquark pair annihilates into a gluon and a photon~\cite{Traxler:1995kx,Steffen:2001pv}. 

Notably, higher-order processes have been shown to play a non-negligible role: Aurenche \textit{et al.} demonstrated that two-loop bremsstrahlung (radiation from charged particles accelerating in strong fields) contributes to photon production at a magnitude comparable to one-loop Compton scattering or annihilation~\cite{Steffen:2001pv,Bhatt:2010cy}. One channel for hard photon emission involving the annihilation of off-shell quarks and antiquarks, accompanied by interactions with additional quarks or gluons-processes that become significant in the dense, high-temperature QGP environment. While early studies of photon production in heavy-ion collisions focused predominantly on Compton scattering and $q\bar{q}$ annihilation, many studies have emphasized the need to incorporate these higher-order contributions to accurately describe experimental observations~\cite{Traxler:1995kx,Steffen:2001pv,Bhatt:2010cy}.

To quantify thermal photon production, we adopt one-loop perturbative QCD calculations supplemented with hard thermal loop (HTL) resummation, a technique that accounts for medium effects (e.g., screening of color charges) in the QGP~\cite{Traxler:1995kx}. This framework yields the following production rates for the dominant processes:

{\it Compton scattering + $q\bar{q}$ annihilation } (C+A):  
   This channel combines contributions from quark-gluon Compton scattering and quark-antiquark annihilation, described by:
   \begin{equation}
   E\frac{dN_{\textrm{C+A}}}{d^{3}pd^{4}x} = \frac{1}{2\pi^{2}}\alpha\alpha_{s}\left(\sum_{f}e_{f}^{2}\right)T^{2}e^{-E/T}\ln\left(\frac{cE}{\alpha_{s}T}\right),
   \label{eq:ca}
   \end{equation}
   where $\alpha = 1/137$ is the electromagnetic fine-structure constant, $\alpha_s$ is the strong coupling constant, and $e_f$ denotes the electric charge of quark flavor $f$ (in units of the electron charge $e$). For this study, we consider light quarks ($u$ and $d$), with $e_u = 2/3$ and $e_d = -1/3$. The constant $c \approx 0.23$ arises from HTL resummation, and the exponential factor $e^{-E/T}$ reflects the Boltzmann distribution of thermal particles in the QGP. The strong coupling is parametrized as a function of temperature~\cite{Karsch:1987kz}:
   \begin{equation}
   \alpha_{s}(T) = \frac{6\pi}{(33-2N_f)\ln(8T/T_c)},
   \end{equation}
   where $N_f = 2$ (number of active quark flavors), $T_c = 0.14$ GeV (QGP-hadron phase transition temperature), and the logarithmic dependence captures the running of $\alpha_s$ with energy scale.

{\it Bremsstrahlung} (Bre) processes:  
   Bremsstrahlung describes photon emission from quarks accelerated by interactions with other partons (quarks or gluons) in the QGP. Its rate is given by:
   \begin{equation}
   E\frac{dN_{\textrm{Bre}}}{d^{3}pd^{4}x} = \frac{1}{8\pi^{5}}\alpha\alpha_{s}\left(\sum_{f}e_{f}^{2}\right)\frac{T^{4}}{E^{2}}e^{-E/T}(J_T - J_L)I(E,T),
   \label{eq:bre}
   \end{equation}
   where $J_T \approx 1.11$ and $J_L \approx -1.06$ are transverse and longitudinal flow parameters, respectively, determined from two loop calculations for two quark flavors and three gluon colors~\cite{Srivastava:1999ekv,Steffen:2001pv}. The function $I(E,T)$ encapsulates kinematic effects via polylogarithmic terms:
   \begin{equation}
   \begin{aligned}
   I(E,T) &= 3\zeta(3) + \frac{\pi^2}{6}\frac{E}{T} + \frac{E^2}{T^2}\ln 2 + 4\textrm{Li}_3(-e^{-|E|/T}) \\
   &+ \frac{2E}{T}\textrm{Li}_2(-e^{-|E|/T}) - \left(\frac{E}{T}\right)^2\ln\left(1 + e^{-|E|/T}\right),
   \end{aligned}
   \end{equation}
   with $\zeta(3) \approx 1.202$ (Riemann zeta function) and $\textrm{Li}_a(z) = \sum_{n=1}^\infty z^n/n^a$ (polylogarithm), which describe the phase space integrals of thermal partons. The $1/E^2$ dependence reflects the soft nature of bremsstrahlung, making it dominant at low photon energies.

{\it $q\bar{q}$-annihilation with additional scattering} (A+S):  
   This process involves quark-antiquark annihilation accompanied by a secondary scattering event in the medium, enhancing photon production at intermediate energies. Its rate is:
   \begin{equation}
   E\frac{dN_{\textrm{A+S}}}{d^{3}pd^{4}x} = \frac{8}{3\pi^{5}}\alpha\alpha_{s}\left(\sum_{f}e_{f}^{2}\right)ETe^{-E/T}(J_T - J_L).
   \label{eq:as}
   \end{equation}
   Unlike bremsstrahlung, this channel scales linearly with photon energy $E$, making it more significant at higher $E$ than the Bremsstrahlung process.

\begin{figure}[tbp!]
\includegraphics[width=0.85\linewidth]{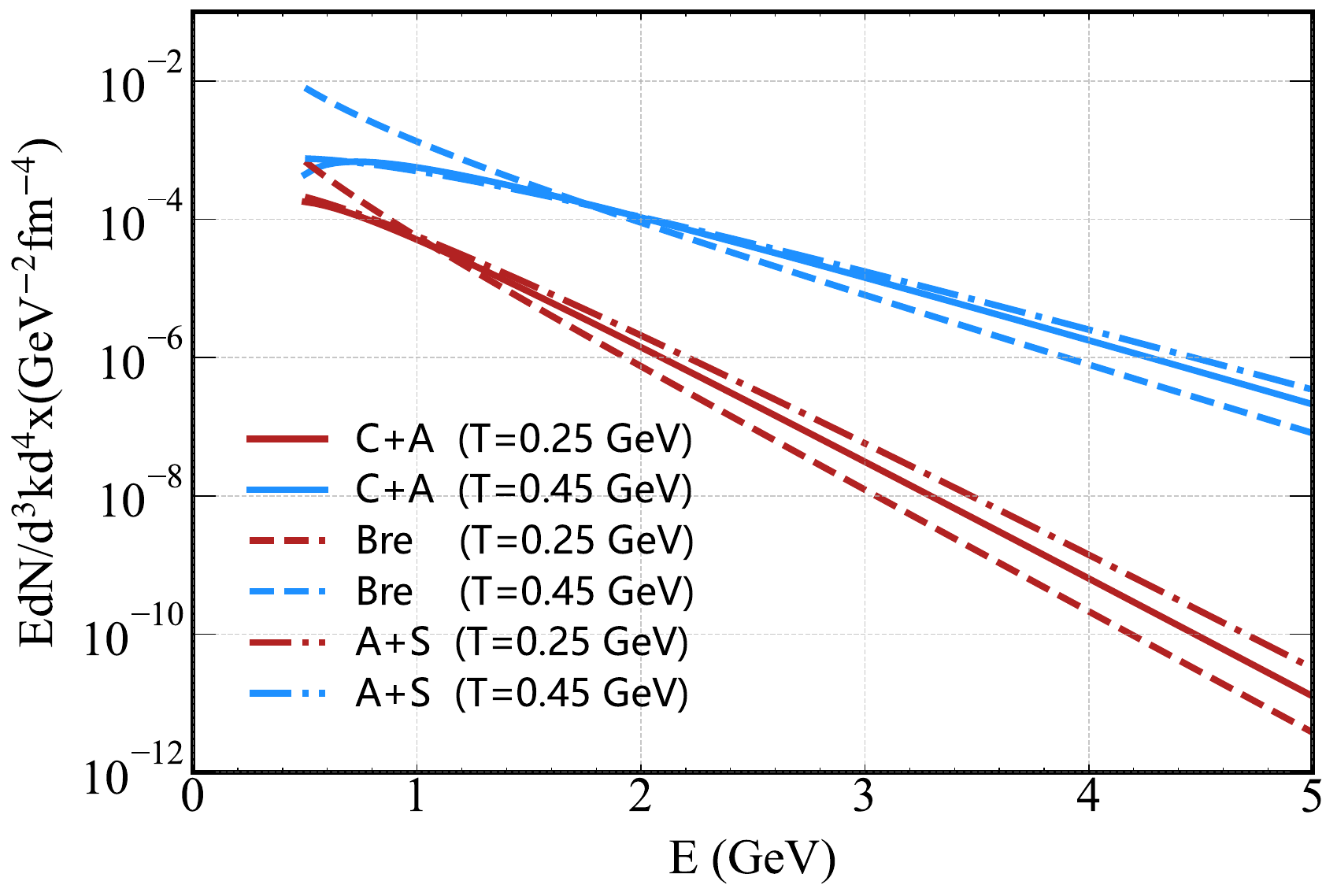} 
\caption{(Color online) Hard thermal photon rates as a function of energy $E$ for fixed temperatures $T=0.25$ GeV and $T=0.45$ GeV, showing contributions from Compton scattering + $q\bar{q}$ annihilation (C+A), bremsstrahlung (Bre), and annihilation with scattering (A+S).}
\label{f:cabsa-1}
\end{figure}

In Fig.~\ref{f:cabsa-1}, we illustrate these three contributions at fixed temperatures $T=0.25$ GeV and $T=0.45$ GeV. We find that Bremsstrahlung dominates the low-energy regime, contributing up to $E \sim 1$ GeV for $T=0.25$ GeV and $E \sim 1.8$ GeV for $T=0.45$ GeV; beyond these energies, the C+A and A+S processes become dominant. This is because Bremsstrahlung is suppressed at high $E$ by the $1/E^2$ factor, while C+A and A+S processes gain strength from their logarithmic or linear dependence on $E$. Additionally, all three rates increase with temperature, a consequence of the higher thermal occupation numbers and more frequent parton interactions in hotter QGP-a trend consistent with previous theoretical studies~\cite{Steffen:2001pv,Bhatt:2010cy}.

The total thermal photon rate is the sum of these individual contributions:
\begin{equation}
E\frac{dN_{\textrm{total}}}{d^{3}pd^{4}x} = E\frac{dN_{\textrm{C+A}}}{d^{3}pd^{4}x} + E\frac{dN_{\textrm{Bre}}}{d^{3}pd^{4}x} + E\frac{dN_{\textrm{A+S}}}{d^{3}pd^{4}x}.
\label{f:cabsa-total}
\end{equation}

\begin{figure}[tbp!]
\includegraphics[width=0.85\linewidth]{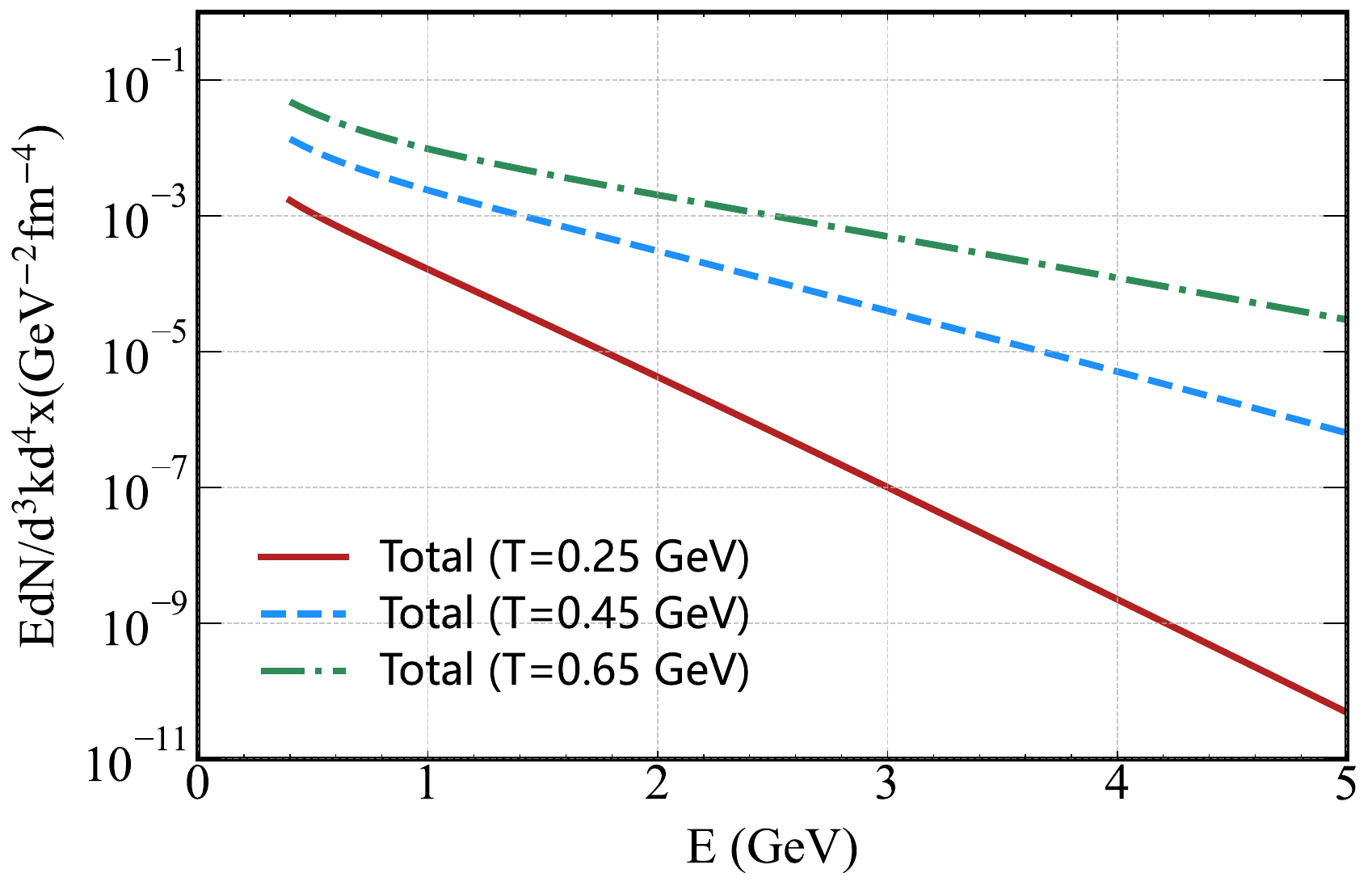} 
\caption{(Color online) Total hard thermal photon rates (Eq.~(\ref{f:cabsa-total})) as a function of energy $E$ for $T=0.25$ GeV, $T=0.45$ GeV, and $T=0.65$ GeV, where ``Total = (C+A) + Bre + (A+S)''.}
\label{f:cabsa-2}
\end{figure}

In Fig.~\ref{f:cabsa-2}, we plot this total rate for $T=0.25$ GeV, $T=0.45$ GeV, and $T=0.65$ GeV, emphasizing the strong temperature dependence of photon production: at $E \sim 3$ GeV, the rate for $T=0.65$ GeV is approximately $10^4$ times larger than that for $T=0.25$ GeV. This enhancement underscores the sensitivity of thermal photon spectra to the QGP's temperature evolution a key motivation for using photons to probe the medium's thermal history.

To connect these differential rates to observable spectra, we integrate the total rate over the QGP's spacetime evolution, following the approach in Refs.~\cite{Traxler:1995kx,Steffen:2001pv,Bhatt:2010cy}:
\begin{equation}
\left(\frac{dN}{d^{2}\pt dy}\right)_{y,~\pt} = Q_c \int_{\tau_0}^{\tau_f} d\tau\,\tau \int_{-y_{\textrm{nuc}}}^{y_{\textrm{nuc}}} d\eta_s \left(E\frac{dN_{\textrm{total}}}{d^{3}pd^{4}x}\right),
\label{eq:photon_int}
\end{equation}
where $\tau_0 = 0.5$ fm/c (initial proper time) and $\tau_f$ is freeze-out proper time, when the QGP cools to $T_c=140$ MeV. The integral over $\tau$ accounts for photon production throughout the QGP's lifetime, weighted by the expansion factor $\tau$ (a consequence of longitudinal boost invariance), while the integral over $\eta_s$ (space-time rapidity) spans the nuclear rapidity range $y_{\textrm{nuc}}$. The factor $Q_c \sim 180$ fm$^2$ is the nuclear cross section for Au+Au collisions~\cite{Srivastava:1999ekv,Traxler:1995kx,Steffen:2001pv,Bhatt:2010cy}.

Within the Victor-Bjorken flow framework (neglecting transverse expansion), the fluid four-velocity is $u^\mu = (\cosh\eta_s, 0, 0, \sinh\eta_s)$, and the photon energy in the QGP rest frame is $E = \pt \cosh(y - \eta_s)$, where $\pt$ is the transverse momentum and $y$ is the photon rapidity. This relation links the photon's observed momentum to its energy in the medium, enabling the transformation from theoretical rates to experimental observables.

It is important to note limitations of the current framework: we focus on ideal magnetohydrodynamic evolution and thus neglect (1) dissipative corrections to the distribution function from viscosity or electromagnetic fields~\cite{Bhatt:2010cy,Sun:2023pil}, which could modify photon rates at late times, and (2) thermal photon production in the hot hadron gas (HHG) phase, which contributes to low-$\pt$ spectra after QGP freeze-out~\cite{Steffen:2001pv,Shen:2013cca,Chatterjee:2017akg}. 
This approach focuses on the macroscopic thermodynamic modulation of the QGP by magnetic fields, which is complementary to quantum field theory (QFT)-based treatments that capture microscopic quantum corrections (e.g., vertex/propagator modifications) to emission rates~\cite{Hattori:2020htm, Wang:2024gnh,Sun:2023pil}; magnetic field-induced modifications to parton interactions, along with these complementary effects, will be explored in future work to provide a more comprehensive description of thermal photon production in heavy-ion collisions.


\section{Results and discussion}  
\label{section-4}
In this section, we investigate the effect of external magnetic field in magnetohydrodynamic evolution and subsequent thermal photon rate spectra.  
Once we get the temperature profile from the magnetohydrodynamics, we can calculate the photon production rates. The total thermal photon spectrum is obtained by adding different rates using the Eqs.~(\ref{eq:ca}), (\ref{eq:bre}), (\ref{eq:as}) and convoluting with the space time evolution of the heavy-ion collision with Eq.~(\ref{eq:photon_int}). The final proper time $\tau_{f}$ is the time where the QGP temperature evolute to the freeze-out temperature $T_{f}$. For simplicity, we following the Refs.~\cite{Traxler:1995kx,Steffen:2001pv,Bhatt:2010cy} and considering the photon production in mid-rapidity region ($y=0$) in all calculations. In the following calculation, we use the values for RHIC experiment given as $T_{0}=0.31$ GeV (initial temperature), $T_{f}=0.14$ GeV, $\tau_{0}=0.5$ fm, and $y_{\textrm{nuc}}=5.3$, which taken from Refs.~\cite{Steffen:2001pv,Bhatt:2010cy}.

\begin{figure}[tbp!]
\includegraphics[width=0.85\linewidth]{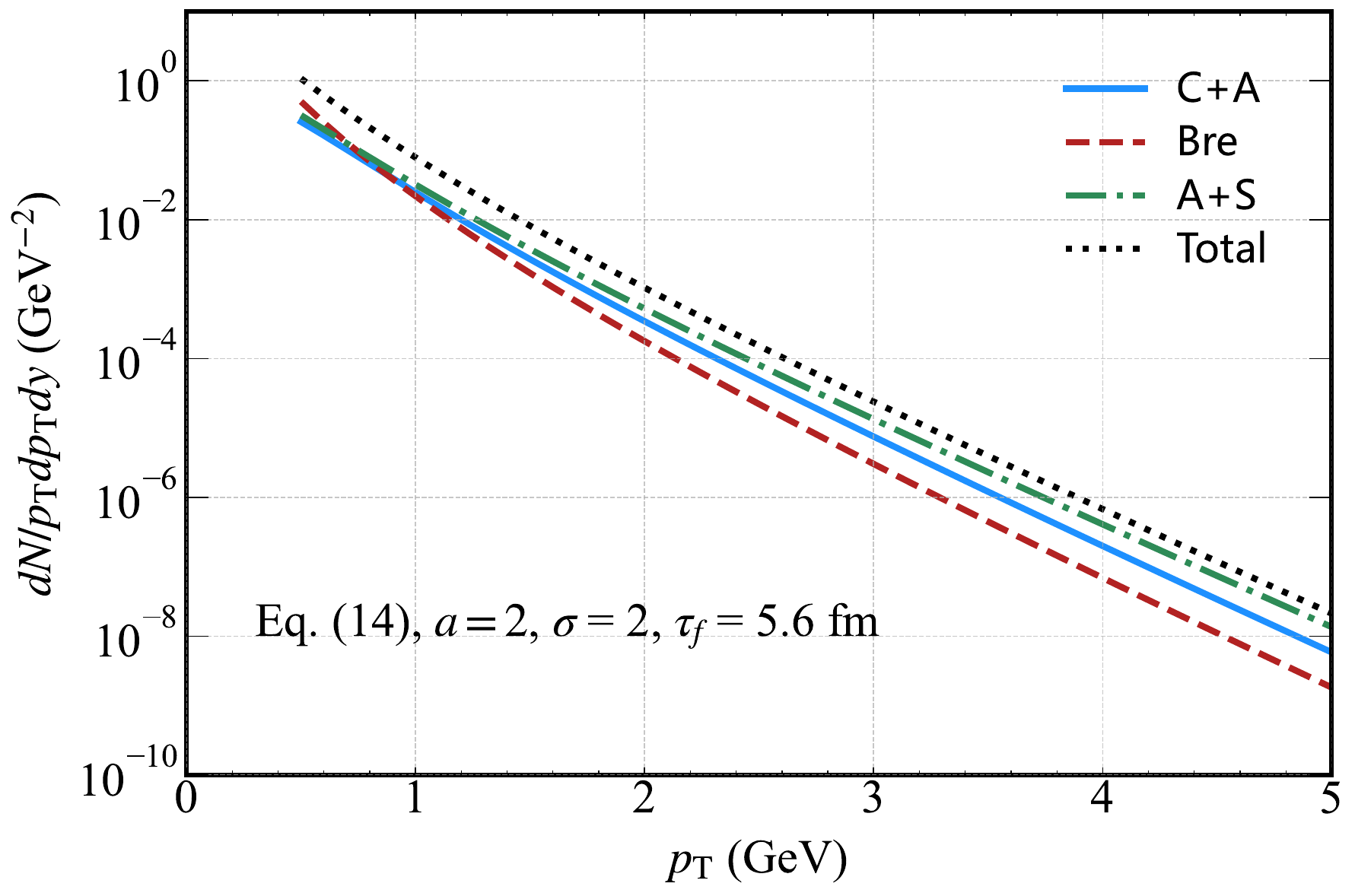} 
\caption{(Color online) Hard thermal photon spectrum from the QGP in the presence of an external magnetic field (with decay parameter \(a=2\) and initial strength parameter \(\sigma=2\)), showing contributions from Compton scattering + \(q\bar{q}\) annihilation (C+A), bremsstrahlung (Bre), annihilation with scattering (A+S), and the total yield. }
\label{f:fig6_total}
\end{figure}

In Fig.~\ref{f:fig6_total}, we present the photon rate calculated from different source. The temperature profiles are obtained from the MHD solutions Eq.~(\ref{T_mhd_1}) with the external magnetic field decay parameter $a=2$ and strength parameter $\sigma=2$. We find that the Bremsstrahlung (Bre) contributes to the photon rate up to $\pt\sim 0.6$ GeV, afterwards the Compton scattering toghter with $q\bar{q}-$annihilation (C+A) and annihilation with scattering processes (A+S) play the dominant role. We further plot the total photon production rate, one finds that the rate close to the A+S distribution at high $\pt$ region.

\begin{figure}[tbp!]
\includegraphics[width=0.85\linewidth]{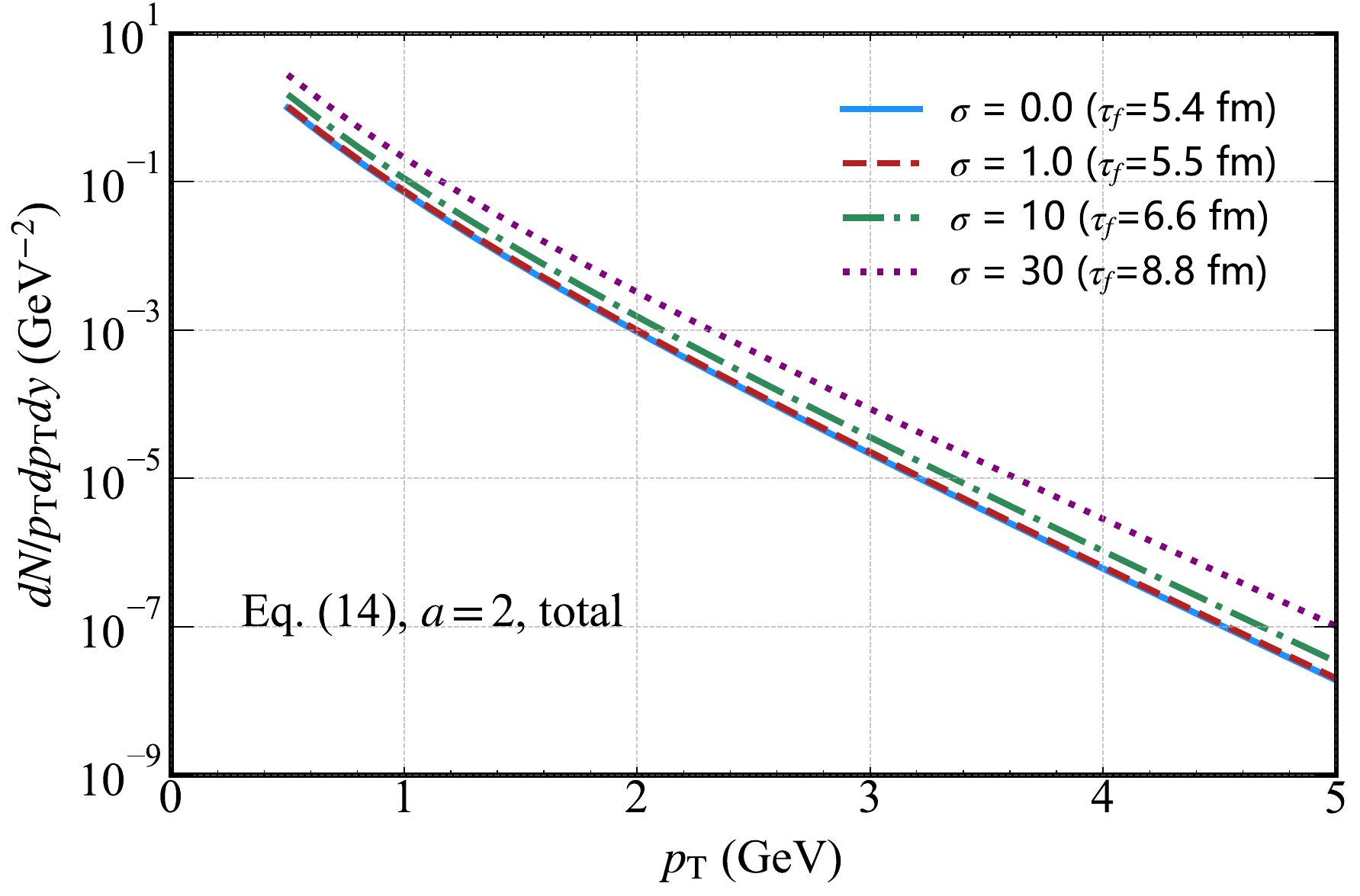} 
\caption{(Color online) Hard thermal photon spectrum from the QGP in the presence of an external magnetic field, showing the total yield for different initial magnetic field strength parameters $\sigma=0.0,~1.0,~10,~30$ with a fixed decay parameter $a=2$ as a function of transverse momentum $\pt$.}
\label{f:fig7_total}
\end{figure}

In Fig.~\ref{f:fig7_total}, we plot the photon rate for initial magnetic field strengths \(\sigma=0,~1,~10,~30\) with the magnetic field decay parameter \(a=2\), where the temperature profiles are taken from the MHD solution given by Eq.~(\ref{T_mhd_1}).
One finds that a strong initial magnetic field (\(\sigma=30\)) leads to a noticeable enhancement in the photon rate across the entire \(\pt\) range. This behavior arises because the external magnetic field exerts a slowing effect on the temperature evolution of the QGP, as clearly shown in Fig.~\ref{f:ideal-mhd}, thereby prolonging the effective time window for thermal photon emission. In contrast, for weak magnetic fields (e.g., \(\sigma=1\)), the influence on the photon rate is negligible. The resulting spectrum is nearly identical to that of the field-free case (\(\sigma=0\)), since the magnetic forces here are too weak to significantly perturb the QGP's thermal dynamics. This observation underscores that the strength of the initial magnetic field plays an important role in shaping the electromagnetic signatures of the QGP.

\begin{figure}[tbp!]
\includegraphics[width=0.85\linewidth]{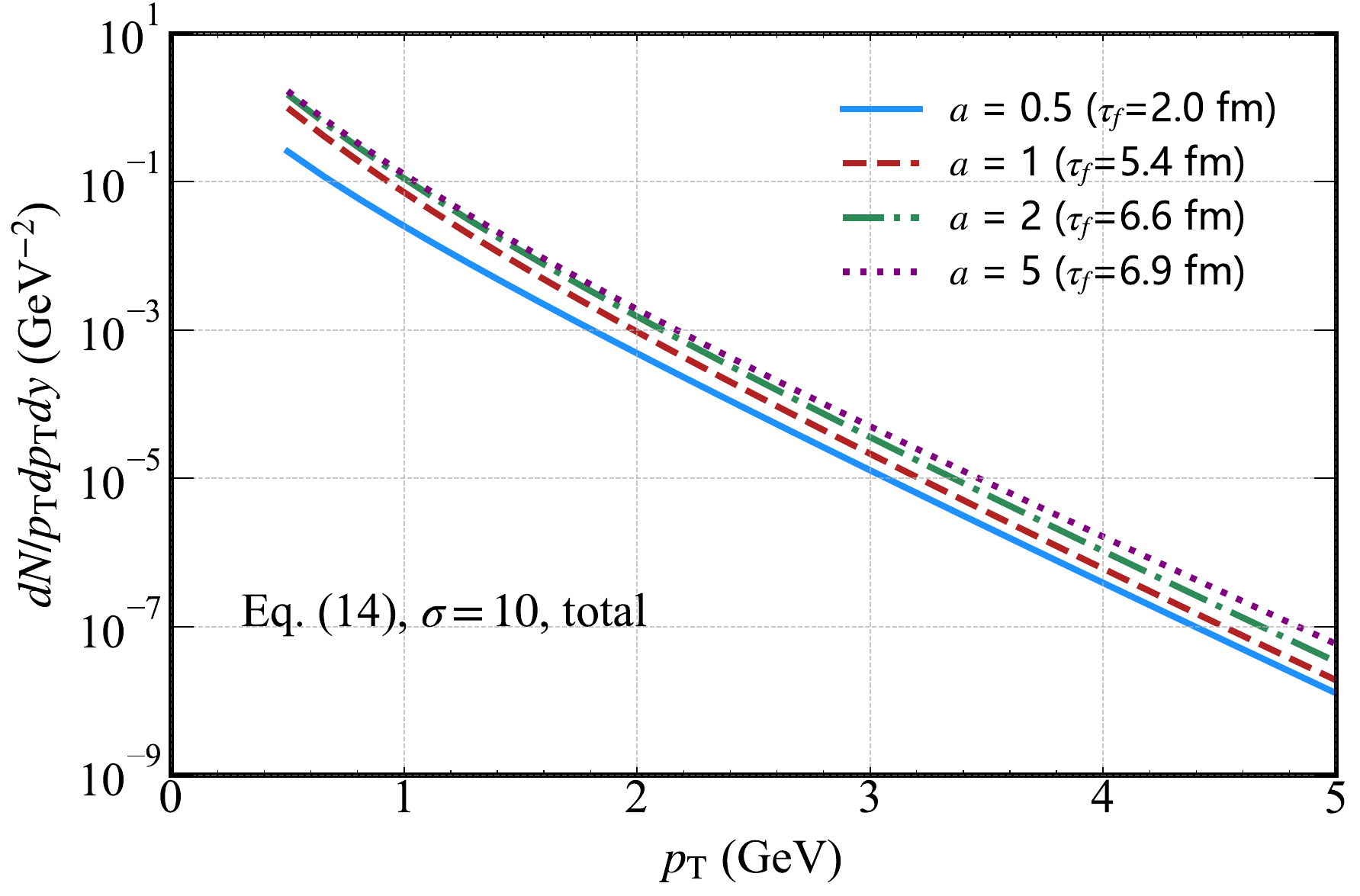} 
\caption{(Color online) Hard thermal photon spectrum from the QGP in the presence of an external magnetic field, showing the total yield for different initial magnetic field decay parameter $a=0.5,~1,~2,~5$ with a fixed strength parameter $\sigma=10$ as a function of transverse momentum $\pt$.}
\label{f:fig8_total}
\end{figure}

In Fig.~\ref{f:fig8_total}, we present the photon spectrum with magnetic field strength parameter $\sigma=10$ and varying decay parameters $a=0.5,~1,~2,~5$. The more pronounced splitting in the large $\pt$ region for $a=1,~2,~5$ can be attributed to the distinct temporal evolution of the magnetic field. Larger decay parameters (e.g., $a=5$) correspond to faster magnetic field decay, which means the magnetic field exerts a more intense influence in the early stages of QGP evolution, when temperatures are higher and hard scattering processes (responsible for large $\pt$ photons) are more active. In contrast, smaller $a$ (slower decay) leads to a more prolonged but weaker magnetic effect. For large $\pt$ photons, which are predominantly produced via high-momentum transfer processes sensitive to the early, high-temperature phase of QGP, the differing timing and intensity of magnetic field effects due to varying $a$ values result in more significant discrepancies in their production rates. This amplifies the splitting observed in the large $\pt$ region for $a=1,~2,~5$. 

\begin{figure}[tbp!]
\includegraphics[width=0.85\linewidth]{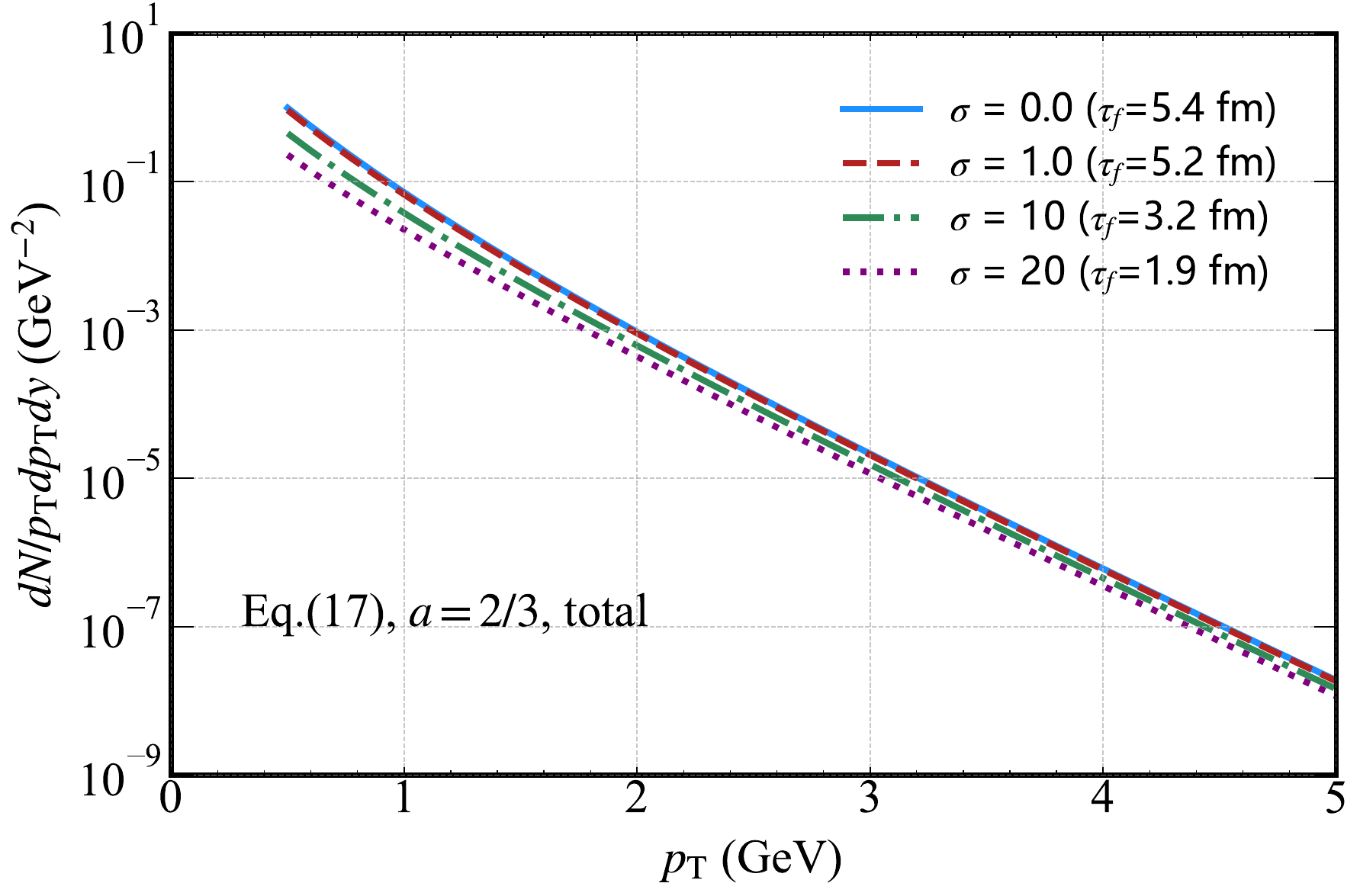}\\
\includegraphics[width=0.85\linewidth]{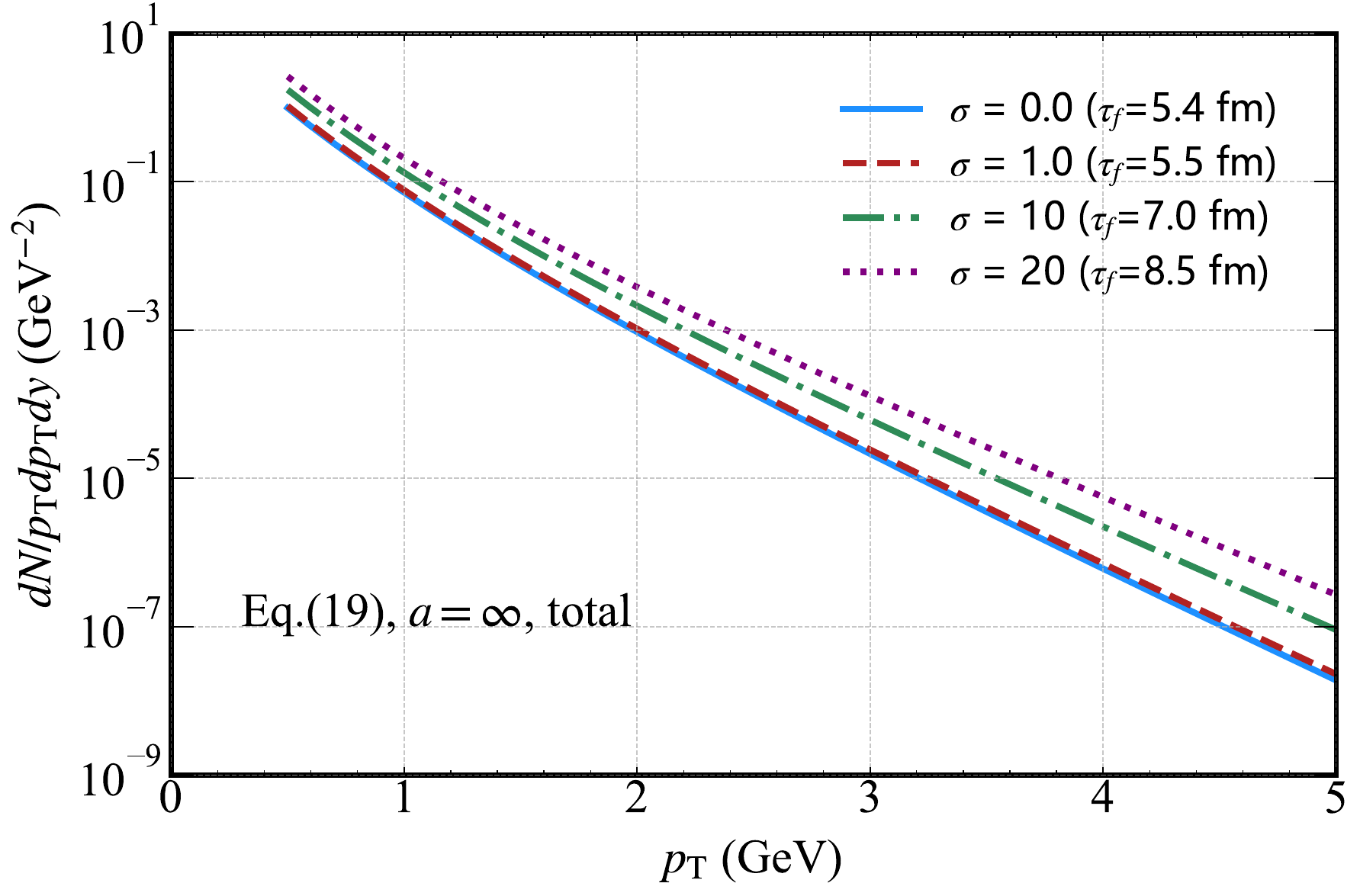}
\caption{(Color online) Hard thermal photon spectra from the QGP in the presence of an external magnetic field. Upper panel: Total yield for different initial magnetic field strength parameters $\sigma=0.0,~1.0,~10,~20$ with magnetic field decay parameter $a=2/3$ (Eq.~(17)) . Lower panel: Total yield for the same $\sigma$ values with magnetic field decay parameter $a=\infty$ (Eq.~(19)).}
\label{f:fig9_total}
\end{figure}

In Fig.~\ref{f:fig9_total} upper panel, we present the photon spectrum for varying initial magnetic field strength parameters $\sigma=0.0,~1.0,~10,~20$ with the magnetic field decay parameter fixed at $a=2/3$. The temperature profiles are derived from the corresponding MHD solutions (e.g., Eq.~(\ref{T_mhd_4}) as applicable). 
A clear trend emerges: as $\sigma$ increases, the photon yield across the $\pt$ spectrum decreases. When magnetic field decay parameter \(a = 2/3\), thermal-photon spectrum variations are governed by initial magnetic field strength \(\sigma\). 
Physically, \(a = 2/3\) lies within \(0 < a \leq \frac{1 + c_s^2}{2}\) (where \(c_s^2\) is the speed of sound), where the fluid expends additional energy density to expel the magnetic field~\cite{Roy:2015kma,Pu:2016ayh}; larger \(\sigma\) requires more energy compensation, accelerating temperature drop. Thus, photon yield for \(a = 2/3\) decreases with increasing \(\sigma\), contrasting with pervious \(a \neq 2/3\). 
This temperature-driven spectrum modulation is a universal feature in Victor-Bjorken-type ideal fluids~\cite{Roy:2015kma,Pu:2016ayh}, accelerated magnetohydrodynamics~\cite{She:2019wdt}, and viscous magnetohydrodynamic systems \cite{Jiang:2024mts}, confirming the physical mechanism's robustness.

In Fig.~\ref{f:fig9_total} lower panel, we show the photon spectrum under the same set of initial magnetic field strength parameters $\sigma=0.0,~1.0,~10,~20$ but with the magnetic field decay parameter $a=\infty$ (Eq.~(\ref{T_mhd_5})). In contrast to the $a=2/3$ case, a reverse trend is observed: larger $\sigma$ values lead to higher photon yields across the $\pt$ spectrum. This is likely due to the decay nature of the magnetic field (as $a=\infty$ implies super fast decay) exerting a enhancing effect on the QGP's thermal and dynamical processes, thereby promoting photon production. The contrasting dependencies of photon yield on $\sigma$ between the two figures highlight the critical role of the magnetic field decay parameter $a$ in mediating the influence of initial magnetic field strength on QGP photon emission.

\begin{figure}[tbp!]
\includegraphics[width=0.85\linewidth]{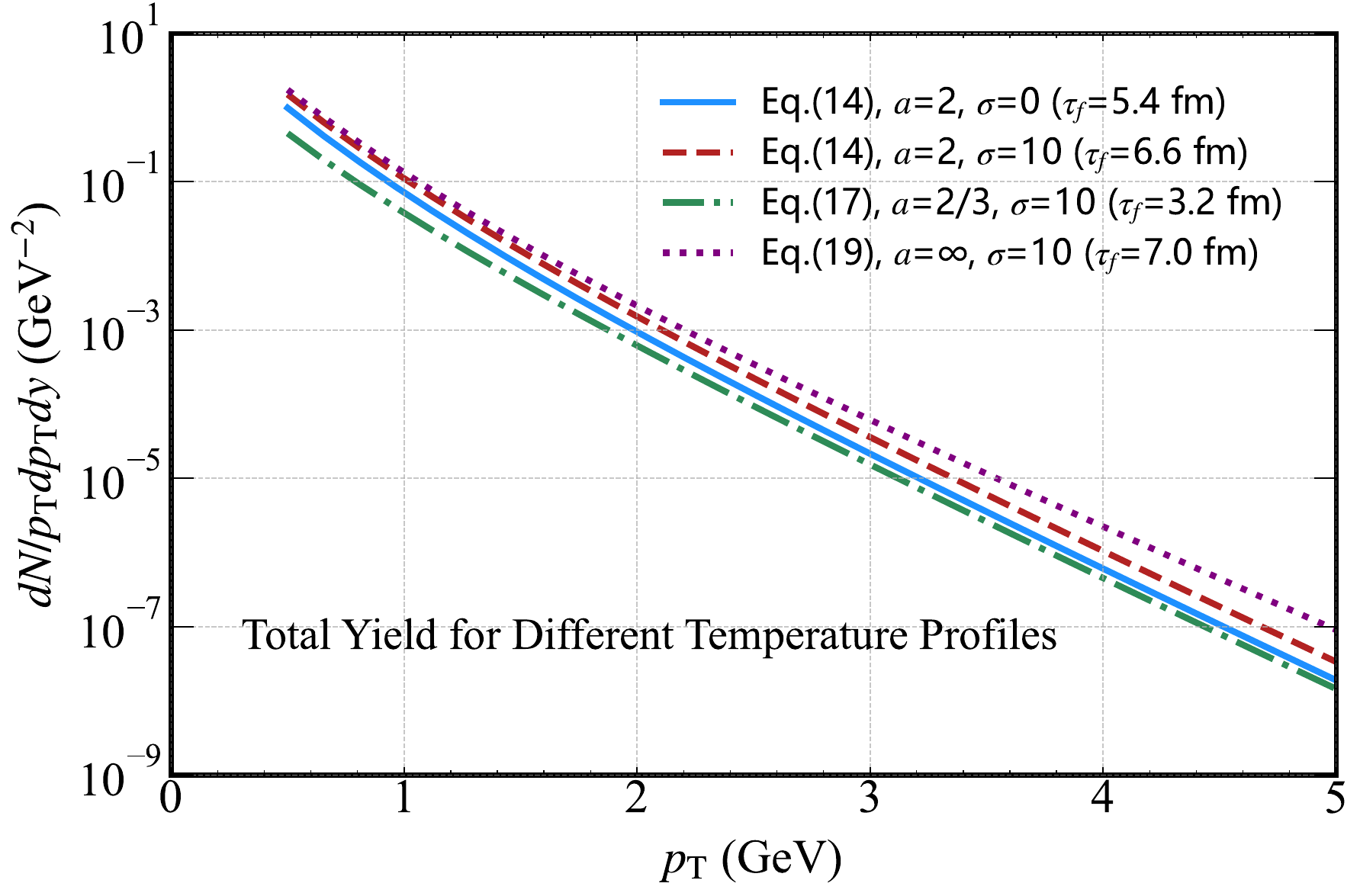}
\caption{(Color online) Hard thermal photon spectrum from the QGP in the presence of an external magnetic field, showing the total yield for different magnetic field decay parameters: \(a=2/3\) (Eq.~(17), \(\sigma=10\)), \(a=2\) (Eq.~(14), \(\sigma=10\) and \(\sigma=0\)), and \(a=\infty\) (Eq.~(19), \(\sigma=10\)).}
\label{f:fig11_total}
\end{figure}

In Fig.~\ref{f:fig11_total}, we present the thermal photon spectra for varying magnetic field decay parameters $a=2/3,~2,~\infty$ with fixed initial magnetic field strength parameter $\sigma=10$, alongside the spectrum for $a=2$ and $\sigma=0$ (ideal MHD case). The temperature profiles are derived from the corresponding MHD solutions (Eqs.~(\ref{T_mhd_1}), (\ref{T_mhd_4}), (\ref{T_mhd_5}) as applicable). A clear trend is observed: as the magnetic field decay parameter $a$ increases, the photon yield across the $\pt$ spectrum rises. This behavior arises because larger $a$ corresponds to faster magnetic field decay, which intensifies the field's influence in the early, high temperature phase of QGP evolution when hard scattering processes (critical for photon production) are most active. Notably, for $a=\infty$ with $\sigma=10$, the photon yield is enhanced compared to the ideal MHD case ($a=2,~\sigma=0$), serving as an upper bound for photon production under magnetic field decay effects. In contrast, for $a=2/3$ with $\sigma=10$, the magnetic field's decay behavior suppresses the photon yield relative to the $\sigma=0$ case. This suppression stems from the slower decay rate of $a=2/3$ perturbing the QGP's thermal dynamics and diminishes emission processes.

\begin{figure}[tbp!]
\includegraphics[width=0.85\linewidth]{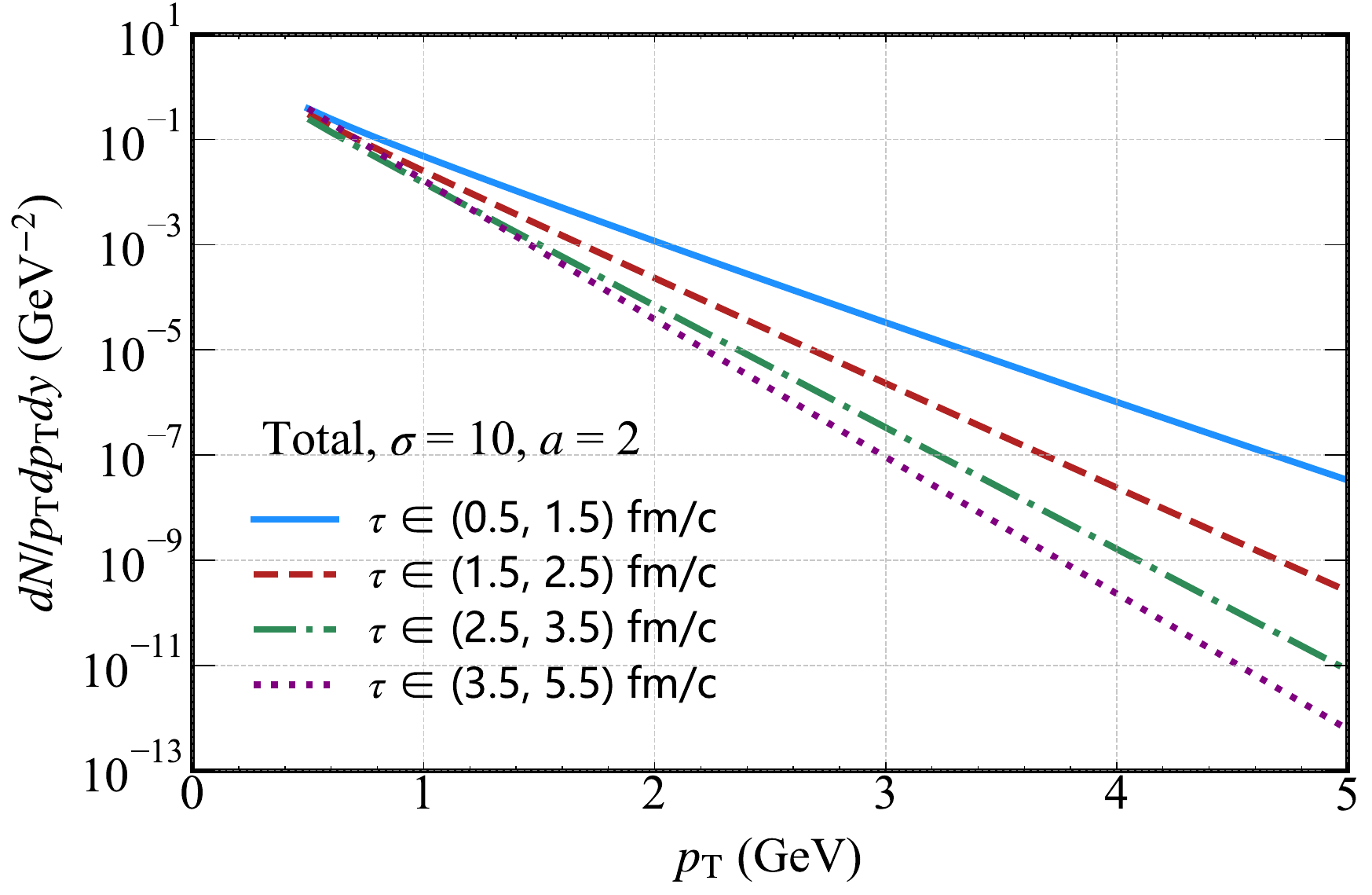}
\caption{(Color online) Hard thermal photon spectrum from the QGP in the presence of an external magnetic field, showing the total yield and contributions from different proper time ($\tau$) intervals.}
\label{f:fig12_total}
\end{figure}

In Fig.~\ref{f:fig12_total}, we show the thermal photon spectra for different proper time intervals [(0.5, 1.5) fm/c, (1.5, 2.5) fm/c, (2.5, 3.5) fm/c, (3.5, 5.5) fm/c] under the condition of magnetic field decay parameter $a=2$ and initial strength parameter $\sigma=10$. The temperature profiles are derived from the corresponding MHD solutions (Eqs.~(\ref{T_mhd_1})). A key observation is that low $\pt$ photons receive contributions from all stages of QGP evolution. This is because low-momentum photons are predominantly produced via relatively low-energy scattering processes, which remain active across the entire lifetime of the QGP as long as the system maintains sufficient temperature. In contrast, high $\pt$ photons are primarily generated in the early stages of QGP evolution (specifically the (0.5, 1.5) fm/c interval). This behavior arises from the fact that high $\pt$ photons require high-momentum transfer processes, which are most efficient during the early, high-temperature phase of the QGP where the collision rate and thermal energy are highest.

\begin{figure}[tbp!]
\includegraphics[width=0.85\linewidth]{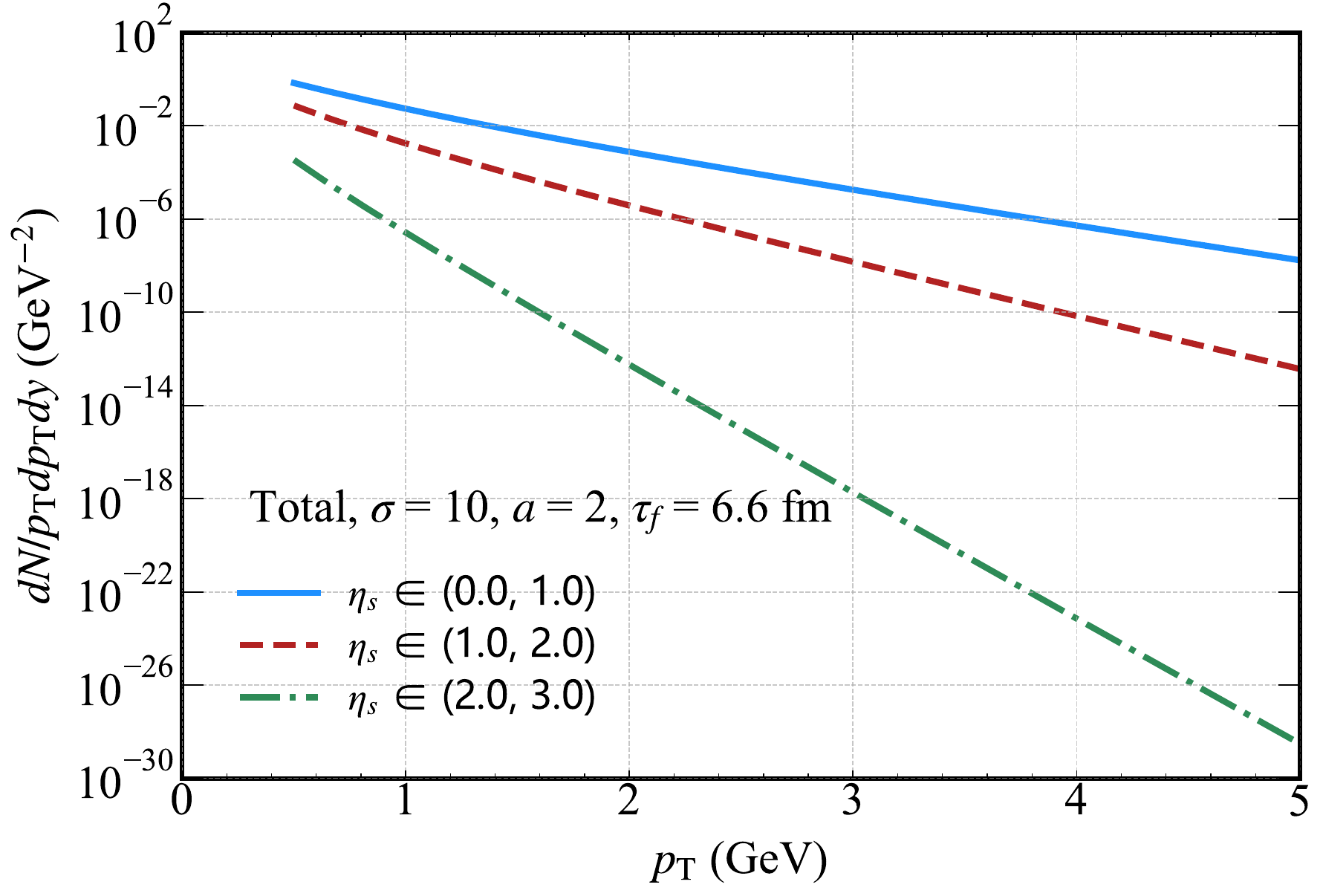}
\caption{(Color online) Hard thermal photon spectrum from the QGP in the presence of an external magnetic field, showing the total yield and contributions from different space-time rapidity ($\eta_{s}$) intervals.}
\label{f:fig13_total}
\end{figure}

Finally, in Fig.~\ref{f:fig13_total}, we plot the thermal photon spectra for different rapidity intervals [$\eta_{s}\in$(0.0, 1.0), $\eta_{s}\in$ (1.0, 2.0), $\eta_{s}\in$ (2.0, 3.0)] under the conditions of magnetic field decay parameter $a=2$ and initial strength parameter $\sigma=10$. A distinct feature observed is that for photons at $y=0$, the yield is predominantly contributed by the central rapidity interval $\eta_{s}\in$(0.0, 1.0). This can be attributed to the fact that particle production in heavy-ion collisions is typically most intense around mid-rapidity, where the density of interacting particles and the corresponding thermal emission processes are most active.  Furthermore, a clear decreasing trend is evident with increasing rapidity: the contribution to the photon yield diminishes as the rapidity interval moves away from the center (e.g., from (0.0, 1.0) to (2.0, 3.0)). This behavior aligns with the expected momentum distribution of particles in the QGP, where the number of particles with large rapidities is smaller due to the kinematic constraints of the collision system. These results highlight the sensitivity of thermal photon production to the rapidity distribution, with central rapidity regions playing a dominant role in shaping the overall yield at $y=0$.

\section{Conclusions}  
\label{section-5}

In this work, we investigate the influence of external magnetic fields on thermal photon production in the quark-gluon plasma (QGP) within a magnetohydrodynamic (MHD) framework. We adopt the relativistic fluid theory to describe the energy-momentum tensor of the QGP in the presence of magnetic fields, simplifying it to the ideal fluid limit under the non-resistance approximation (Victor-Bjorken type). The magnetic field is assumed to decay via a power-law in proper time, $\overrightarrow{B}(\tau)=\overrightarrow{B}_{0}(\tau_{0}/\tau)^{a}$, where $a$ is the decay parameter and $\sigma$ characterizes the initial field strength. Using the Victor-Bjorken flow assumption~\cite{Roy:2015kma,Pu:2016ayh,Jiang:2024mts}, we revisit the analytical solutions for the QGP temperature evolution (Eq.~(\ref{T_mhd_1})), which shows distinct behaviors depending on $a$ and $\sigma$. For example, larger $a$ or $\sigma$ slows temperature decay, while $a=2/3$ accelerates it for increasing $\sigma$~\cite{Roy:2015kma}.

Thermal photon production rates are calculated by considering three dominant processes~\cite{Traxler:1995kx,Steffen:2001pv,Bhatt:2010cy}: Compton scattering with $q\bar{q}$ annihilation (C+A), bremsstrahlung (Bre), and $q\bar{q}$ annihilation with additional scattering (A+S). The total photon spectrum is obtained by integrating these rates over the QGP's spacetime evolution (Eq.~(\ref{eq:photon_int})), accounting for the temperature profiles from MHD simulations. This approach captures the contributions of different processes across momentum ranges, with Bremsstrahlung dominating low $\pt$ and C+A/A+S dominating high $\pt$.

Our results reveal that the external magnetic field affect the photon spectrum: (i) Increasing the decay parameter $a$ enhances photon yield across all $\pt$, with $a=\infty$ (super-fast decay) providing an upper bound due to intensified early-stage QGP heating. (ii) For $a=2/3$, larger $\sigma$ suppresses yields by accelerating temperature decay, while for $a=\infty$, larger $\sigma$ enhances yields via prolonged thermal emission. (iii) Low $\pt$ photons receive contributions from all QGP stages, whereas high $\pt$ photons are dominated by early-time production. (iv) Central rapidity intervals (0.0, 1.0) dominate the yield at $y=0$, with contributions diminishing for larger rapidities. These findings highlight the critical role of magnetic field dynamics in QGP electromagnetic signatures, offering insights for heavy-ion collision experiments.

This study, being simple and easily reproducible, first extends studies of photon yields from viscous fluids~\cite{Steffen:2001pv,Bhatt:2010cy}, Gubser flow~\cite{Naik:2025pjt}, and longitudinally accelerated fluids~\cite{Kasza:2025wot} to the magnetohydrodynamic regime. It holds many theoretical potential for further exploration, including photon yields in magnetohydrodynamics with magnetic susceptibility~\cite{Pu:2016ayh}, viscous magnetohydrodynamics~\cite{Jiang:2024mts}, longitudinally accelerated magnetohydrodynamics~\cite{She:2019wdt}, spin-hydrodynamics~\cite{Wang:2021wqq}, and non-extensive (magneto-)hydrodynamics~\cite{Shen:2017pyo}. Such investigations will deepen our understanding of how strong magnetic fields and spin effects influence photon (or dilepton) yields in relativistic hydrodynamic framework, while providing a testbed for thermal photons (or dilepton~\cite{Dwibedi:2025xho,Wu:2024vyc}) from numerical magnetohydrodynamic simulations.

\begin{acknowledgements}
This work was supported by the National Natural Science Foundation of China (NSFC) under Grant Nos.~12305138. Duan She's research is funded by the Startup Research Fund of Henan Academy of Sciences (No. 231820058) and the 2024 Henan Province International Science and Technology Cooperation Projects (No. 242102521068).
\end{acknowledgements}

\bibliographystyle{unsrt}
\bibliography{clv3}

\begin{thebibliography}{10}

\bibitem{STAR:2005gfr}
J.~Adams and others (STAR~Collaboration).
\newblock {\em Nucl. Phys. A}, 757:102--183, 2005.

\bibitem{ALICE:2008ngc}
K.~Aamodt and others (ALICE~Collaboration).
\newblock {\em JINST}, 3:S08002, 2008.

\bibitem{Deng:2012pc}
W.-T. Deng and X.-G. Huang.
\newblock {\em Phys. Rev. C}, 85:044907, 2012.

\bibitem{Li:2016tel}
H.~Li, X.-L. Sheng, and Q.~Wang.
\newblock {\em Phys. Rev. C}, 94(4):044903, 2016.

\bibitem{Gursoy:2014aka}
U.~Gursoy, D.~E. Kharzeev, and K.~Rajagopal.
\newblock {\em Phys. Rev. C}, 89(5):054905, 2014.

\bibitem{Das:2016cwd}
Santosh~K. Das, S.~Plumari, S.~Chatterjee, J.~Alam, F.~Scardina, and V.~Greco.
\newblock {\em Phys. Lett. B}, 768:260--264, 2017.

\bibitem{Gursoy:2018yai}
U.~G\"ursoy, D.~E. Kharzeev, E.~Marcus, K.~Rajagopal, and C.~Shen.
\newblock {\em Phys. Rev. C}, 98(5):055201, 2018.

\bibitem{Chatterjee:2018lsx}
S.~Chatterjee and P.~Bozek.
\newblock {\em Phys. Lett. B}, 798:134955, 2019.

\bibitem{Oliva:2020doe}
L.~Oliva, S.~Plumari, and V.~Greco.
\newblock {\em JHEP}, 05:034, 2021.

\bibitem{Sun:2020wkg}
Y.~Sun, S.~Plumari, and V.~Greco.
\newblock {\em Phys. Lett. B}, 816:136271, 2021.

\bibitem{Jiang:2022uoe}
Z.-F. Jiang, S.-S. Cao, W.-J. Xing, X.-Y. Wu, C.~B. Yang, and B.-W. Zhang.
\newblock {\em Phys. Rev. C}, 105(5):054907, 2022.

\bibitem{Huang:2017tsq}
A.~Huang, Y.~Jiang, S.~Shi, J.~Liao, and P.~Zhuang.
\newblock {\em Phys. Lett. B}, 777:177--183, 2018.

\bibitem{Kharzeev:2007jp}
D.~E. Kharzeev, L.~D. McLerran, and H.~J. Warringa.
\newblock {\em Nucl. Phys. A}, 803:227--253, 2008.

\bibitem{Fukushima:2008xe}
K.~Fukushima, D.~E. Kharzeev, and H.~J. Warringa.
\newblock {\em Phys. Rev. D}, 78:074033, 2008.

\bibitem{Kharzeev:2015znc}
D.~E. Kharzeev, J.~Liao, S.~A. Voloshin, and G.~Wang.
\newblock {\em Prog. Part. Nucl. Phys.}, 88:1--28, 2016.

\bibitem{Kharzeev:2024zzm}
D.~E. Kharzeev, J.~Liao, and P.~Tribedy.
\newblock {\em Int. J. Mod. Phys. E}, 33(09):2430007, 2024.

\bibitem{Liang:2020sgr}
G.-R. Liang, J.~Liao, S.~Lin, L.~Yan, and M.~Li.
\newblock {\em Chin. Phys. C}, 44(9):094103, 2020.

\bibitem{Jiang:2016wve}
Y.~Jiang, S.~Shi, Y.~Yin, and J.~Liao.
\newblock {\em Chin. Phys. C}, 42(1):011001, 2018.

\bibitem{Choudhury:2021jwd}
S.~Choudhury and others (STAR~Collaboration).
\newblock {\em Chin. Phys. C}, 46(1):014101, 2022.

\bibitem{Tang:2019pbl}
A.~H. Tang.
\newblock {\em Chin. Phys. C}, 44(5):054101, 2020.

\bibitem{Gao:2020vbh}
J.-H. Gao, G.-L. Ma, S.~Pu, and Q.~Wang.
\newblock {\em Nucl. Sci. Tech.}, 31(9):90, 2020.

\bibitem{Shi:2017cpu}
S.~Shi, Y.~Jiang, E.~Lilleskov, and J.~Liao.
\newblock {\em Annals Phys.}, 394:50--72, 2018.

\bibitem{Huang:2013iia}
X.-G. Huang and J.~Liao.
\newblock {\em Phys. Rev. Lett.}, 110(23):232302, 2013.

\bibitem{Pu:2014cwa}
S.~Pu, S.-Y. Wu, and D.-L. Yang.
\newblock {\em Phys. Rev. D}, 89(8):085024, 2014.

\bibitem{Kharzeev:2010gd}
D.~E. Kharzeev and H.-U. Yee.
\newblock {\em Phys. Rev. D}, 83:085007, 2011.

\bibitem{STAR:2015wza}
L.~Adamczyk and others (STAR~Collaboration).
\newblock {\em Phys. Rev. Lett.}, 114(25):252302, 2015.

\bibitem{CMS:2017lrw}
A.~M. Sirunyan and others (CMS~Collaboration).
\newblock {\em Phys. Rev. C}, 97(4):044912, 2018.

\bibitem{STAR:2021mii}
M.~Abdallah and others (STAR~Collaboration).
\newblock {\em Phys. Rev. C}, 105(1):014901, 2022.

\bibitem{STAR:2021pwb}
M.~S. Abdallah and others (STAR~Collaboration).
\newblock {\em Phys. Rev. Lett.}, 128(9):092301, 2022.

\bibitem{STAR:2019clv}
J.~Adam and others (STAR~Collaboration).
\newblock {\em Phys. Rev. Lett.}, 123(16):162301, 2019.

\bibitem{ALICE:2019sgg}
S.~Acharya and others (ALICE~Collaboration).
\newblock {\em Phys. Rev. Lett.}, 125(2):022301, 2020.

\bibitem{STAR:2023jdd}
M.~I. Abdulhamid and others (STAR~Collaboration).
\newblock {\em Phys. Rev. X}, 14(1):011028, 2024.

\bibitem{Heinz:2013th}
U.~Heinz and R.~Snellings.
\newblock {\em Ann. Rev. Nucl. Part. Sci.}, 63:123--151, 2013.

\bibitem{Gale:2013da}
C.~Gale, S.~Jeon, and B.~Schenke.
\newblock {\em Int. J. Mod. Phys. A}, 28:1340011, 2013.

\bibitem{Karpenko:2013wva}
Iu. Karpenko, P.~Huovinen, and M.~Bleicher.
\newblock {\em Comput. Phys. Commun.}, 185:3016--3027, 2014.

\bibitem{Jeon:2015dfa}
S.~Jeon and U.~Heinz.
\newblock {\em Int. J. Mod. Phys. E}, 24(10):1530010, 2015.

\bibitem{Becattini:2017gcx}
F.~Becattini and Iu. Karpenko.
\newblock {\em Phys. Rev. Lett.}, 120(1):012302, 2018.

\bibitem{Becattini:2020ngo}
F.~Becattini and M.~A. Lisa.
\newblock {\em Ann. Rev. Nucl. Part. Sci.}, 70:395--423, 2020.

\bibitem{Ze-Fang:2017ppe}
Z.-F. Jiang, C.-B. Yang, M.~Csanad, and T.~Csorgo.
\newblock {\em Phys. Rev. C}, 97(6):064906, 2018.

\bibitem{Zhao:2022ayk}
W.~Zhao, C.~Shen, and B.~Schenke.
\newblock {\em Phys. Rev. Lett.}, 129(25):252302, 2022.

\bibitem{Zhao:2020wcd}
W.~Zhao, C.-M. Ko, Y.-X. Liu, G.-Y. Qin, and H.~Song.
\newblock {\em Phys. Rev. Lett.}, 125(7):072301, 2020.

\bibitem{Jiang:2021ajc}
Z.-F. Jiang, S.-S. Cao, X.-Y. Wu, C.~B. Yang, and B.-W. Zhang.
\newblock {\em Phys. Rev. C}, 105(3):034901, 2022.

\bibitem{Wu:2021fjf}
X.-Y. Wu, G.-Y. Qin, L.-G. Pang, and X.-N. Wang.
\newblock {\em Phys. Rev. C}, 105(3):034909, 2022.

\bibitem{Jiang:2024ekh}
Z.-F. Jiang, S.-S. Cao, and B.-W. Zhang.
\newblock {\em Phys. Rev. C}, 111(3):034906, 2025.

\bibitem{Inghirami:2016iru}
G.~Inghirami, L.~D.~Zanna, A.~Beraudo, M.~Moghaddam, F.~Becattini, and M.~Bleicher.
\newblock {\em Eur. Phys. J. C}, 76(12):659, 2016.

\bibitem{Nakamura:2022idq}
K.~Nakamura, T.~Miyoshi, C.~Nonaka, and H.~R. Takahashi.
\newblock {\em Phys. Rev. C}, 107(1):014901, 2023.

\bibitem{Mayer:2024kkv}
M.~Mayer, A.~Dash, G.~Inghirami, H.~Elfner, L.~Rezzolla, and D.~H. Rischke.
\newblock {\em Phys. Rev. C}, 111(4):044908, 2025.

\bibitem{Ding:2010ga}
H.~T. Ding, A.~Francis, O.~Kaczmarek, F.~Karsch, E.~Laermann, and W.~Soeldner.
\newblock {\em Phys. Rev. D}, 83:034504, 2011.

\bibitem{Ding:2016hua}
H.-T. Ding, O.~Kaczmarek, and F.~Meyer.
\newblock {\em Phys. Rev. D}, 94(3):034504, 2016.

\bibitem{Bali:2013owa}
G.~S. Bali, F.~Bruckmann, G.~Endrodi, and A.~Schafer.
\newblock {\em Phys. Rev. Lett.}, 112:042301, 2014.

\bibitem{Pang:2016yuh}
L.-G. Pang, G.~Endr\H{o}di, and H.~Petersen.
\newblock {\em Phys. Rev. C}, 93(4):044919, 2016.

\bibitem{Jiang:2024bez}
Z.-F. Jiang, Z.-H. Zhang, X.-F. Yuan, and B.-W. Zhang.
\newblock {\em Phys. Rev. C}, 110(1):014902, 2024.

\bibitem{Huang:2024aob}
A.~Huang, X.-Y. Wu, and M.~Huang.
\newblock {\em Phys. Rev. D}, 110(9):094032, 2024.

\bibitem{Fukushima:2024tkz}
K.~Fukushima, K.~Hattori, and K.~Mameda.
\newblock {\em arXiv:~2409.18652}, 9 2024.

\bibitem{Roy:2015kma}
V.~Roy, S.~Pu, L.~Rezzolla, and D.~Rischke.
\newblock {\em Phys. Lett. B}, 750:45--52, 2015.

\bibitem{Pu:2016ayh}
S.~Pu, V.~Roy, L.~Rezzolla, and D.~H. Rischke.
\newblock {\em Phys. Rev. D}, 93(7):074022, 2016.

\bibitem{Siddique:2019gqh}
I.~Siddique, R.-j. Wang, S.~Pu, and Q.~Wang.
\newblock {\em Phys. Rev. D}, 99(11):114029, 2019.

\bibitem{Peng:2022cya}
H.-H. Peng, S.~Wu, R.-j. Wang, D.~She, and S.~Pu.
\newblock {\em Phys. Rev. D}, 107(9):096010, 2023.

\bibitem{She:2019wdt}
D.~She, Z.-F. Jiang, D.~Hou, and C.-B. Yang.
\newblock {\em Phys. Rev. D}, 100(11):116014, 2019.

\bibitem{HaddadiMoghaddam:2020ihi}
M.~Haddadi~M., W.~M. Alberico, D.~She, A.~F. Kord, and B.~Azadegan.
\newblock {\em Phys. Rev. D}, 102(1):014017, 2020.

\bibitem{Shokri:2018qcu}
M.~Shokri and N.~Sadooghi.
\newblock {\em JHEP}, 11:181, 2018.

\bibitem{Biswas:2020rps}
R.~Biswas, A.~Dash, N.~Haque, S.~Pu, and V.~Roy.
\newblock {\em JHEP}, 10:171, 2020.

\bibitem{Cordeiro:2023ljz}
I.~Cordeiro, E.~Speranza, K.~Ingles, F.~Bemfica, and J.~Noronha.
\newblock {\em Phys. Rev. Lett.}, 133(9):091401, 2024.

\bibitem{Fang:2024skm}
Z.~Fang, K.~Hattori, and J.~Hu.
\newblock {\em Phys. Rev. D}, 110(5):056049, 2024.

\bibitem{Rocha:2023ilf}
G.~S. Rocha, D.~Wagner, G.~S. Denicol, J.~Noronha, and D.~H. Rischke.
\newblock {\em Entropy}, 26(3):189, 2024.

\bibitem{Most:2021rhr}
E.~R. Most and J.~Noronha.
\newblock {\em Phys. Rev. D}, 104(10):103028, 2021.

\bibitem{Bhatt:2010cy}
J.~R. Bhatt, H.~Mishra, and V.~Sreekanth.
\newblock {\em JHEP}, 11:106, 2010.

\bibitem{Gale:2003iz}
C.~Gale and Kevin~L. Haglin.
\newblock pages 364--429, 6 2003.

\bibitem{Gale:2009gc}
C.~Gale.
\newblock {\em Landolt-Bornstein}, 23:445, 2010.

\bibitem{dEnterria:2005jvq}
D.~G. d'Enterria and D.~Peressounko.
\newblock {\em Eur. Phys. J. C}, 46:451--464, 2006.

\bibitem{Linnyk:2013hta}
O.~Linnyk, V.~P. Konchakovski, W.~Cassing, and E.~L. Bratkovskaya.
\newblock {\em Phys. Rev. C}, 88:034904, 2013.

\bibitem{Wang:2020dsr}
I.~A. Wang, X.and~Shovkovy, L.~Yu, and M.~Huang.
\newblock {\em Phys. Rev. D}, 102(7):076010, 2020.

\bibitem{Steffen:2001pv}
F.~D. Steffen and M.~H. Thoma.
\newblock {\em Phys. Lett. B}, 510:98--106, 2001.
\newblock [Erratum: Phys.Lett.B 660, 604--606 (2008)].

\bibitem{Naik:2025pjt}
Lakshmi~J. Naik and V.~Sreekanth.
\newblock {\em Eur. Phys. J. C}, 85(6):664, 2025.

\bibitem{Wang:2021eud}
X.~Wang and I.~Shovkovy.
\newblock {\em Eur. Phys. J. C}, 81(10):901, 2021.

\bibitem{Shen:2013cca}
C.~Shen, U.~W. Heinz, J.-F. Paquet, I.~Kozlov, and C.~Gale.
\newblock {\em Phys. Rev. C}, 91(2):024908, 2015.

\bibitem{David:2019wpt}
G.~David.
\newblock {\em Rept. Prog. Phys.}, 83(4):046301, 2020.

\bibitem{Schenke:2006yp}
B.~Schenke and M.~Strickland.
\newblock {\em Phys. Rev. D}, 76:025023, 2007.

\bibitem{Wu:2024vyc}
X.-Y. Wu, H.~Gao, B.~Forster, C.~Gale, and S.~Jackson, G.and~Jeon.
\newblock {\em Phys. Rev. Lett.}, 134(24):242301, 2025.

\bibitem{Chatterjee:2017akg}
R.~Chatterjee, P.~Dasgupta, and D.~K. Srivastava.
\newblock {\em Phys. Rev. C}, 96(1):014911, 2017.

\bibitem{Vovchenko:2016ijt}
V.~Vovchenko, Iu.~A. Karpenko, M.~I. Gorenstein, L.~M. Satarov, I.~N. Mishustin, B.~K{\"a}mpfer, and H.~Stoecker.
\newblock {\em Phys. Rev. C}, 94(2):024906, 2016.

\bibitem{Traxler:1995kx}
C.~T. Traxler and M.~H. Thoma.
\newblock {\em Phys. Rev. C}, 53:1348--1352, 1996.

\bibitem{Kasmaei:2019ofu}
B.~S. Kasmaei and M.~Strickland.
\newblock {\em Phys. Rev. D}, 102(1):014037, 2020.

\bibitem{Jiang:2024mts}
Z.-F. Jiang, S.-Y. Liu, T.-Y. Hu, H.-J. Zheng, and D.~She.
\newblock {\em Chin. Phys.}, 49(11):114104, 2025.

\bibitem{Kasza:2025wot}
G.~L. Kasza.
\newblock {\em arXiv: 2507.13240}, 7 2025.

\bibitem{Wang:2024gnh}
X.~Wang and I.~A. Shovkovy.
\newblock {\em Phys. Rev. D}, 110(11):116005, 2024.

\bibitem{Denicol:2018rbw}
G.~S. Denicol, X.-G. Huang, E.~Moln\'ar, G.~M. Monteiro, H.~Niemi, J.~Noronha, D.~H. Rischke, and Q.~Wang.
\newblock {\em Phys. Rev. D}, 98(7):076009, 2018.

\bibitem{Muronga:2001zk}
A.~Muronga.
\newblock {\em Phys. Rev. Lett.}, 88:062302, 2002.
\newblock [Erratum: Phys.Rev.Lett. 89, 159901 (2002)].

\bibitem{Muronga:2003ta}
A.~Muronga.
\newblock {\em Phys. Rev. C}, 69:034903, 2004.

\bibitem{Karsch:1987kz}
F.~Karsch.
\newblock {\em Z. Phys. C}, 38:147, 1988.

\bibitem{Srivastava:1999ekv}
D.~K. Srivastava.
\newblock {\em Eur. Phys. J. C}, 10:487--490, 1999.
\newblock [Erratum: Eur.Phys.J.C 20, 399--400 (2001)].

\bibitem{Sun:2023pil}
J.-A. Sun and L.~Yan.
\newblock {\em Phys. Lett. B}, 858:139046, 2024.

\bibitem{Hattori:2020htm}
K.~Hattori, H.~Taya, and S.~Yoshida.
\newblock {\em JHEP}, 01:093, 2021.

\bibitem{Wang:2021wqq}
D.-L. Wang, X.-Q. Xie, S.~Fang, and S.~Pu.
\newblock {\em Phys. Rev. D}, 105(11):114050, 2022.

\bibitem{Shen:2017pyo}
K.~Shen, T.~S. Biro, and E.~Wang.
\newblock {\em Physica A}, 492:2353--2360, 2018.

\bibitem{Dwibedi:2025xho}
A.~Dwibedi, A.~K. Panda, S.~Ghosh, and V.~Roy.
\newblock {\em arXiv: 2508.16988}, 8 2025.

\end{thebibliography}

\end{document}